\documentclass[twocolumn]{aastex631}

\usepackage{multirow}
\usepackage{amsmath}

\newcommand\obj{I\,Zw\,1}

\newcommand\softw{\textsc}

\newcommand\telescope{\textit}

\newcommand\Msun{\ifmmode {$M$_{\odot}}\else{M$_{\odot}$}\fi}

\received{XXX}
\revised{XXX}
\accepted{XXX}

\submitjournal{ApJ}

\shorttitle{Jets in the super-Eddington AGN \obj}
\shortauthors{Yang et al.}
\begin{document}

%\title{Template \aastex Article with Examples: 
%v6.31\footnote{Released on March, 1st, 2021}}

%\title{Jet and wind from a rapidly growing supermassive black hole}
\title{Unveiling the small-scale jets in the rapidly growing supermassive black hole \obj}

\correspondingauthor{Xiaolong Yang}
\email{yangxl@shao.ac.cn}

\author[0000-0002-4439-5580]{Xiaolong Yang}
\affiliation{Shanghai Astronomical Observatory, Chinese Academy of Sciences, Shanghai 200030, China}
\affiliation{Kavli Institute for Astronomy and Astrophysics, Peking University, Beijing 100871, China}
\affiliation{Shanghai Key Laboratory of Space Navigation and Positioning Techniques, Shanghai 200030, China}

\author[0000-0002-9728-1552]{Su Yao}
\affiliation{Max-Planck-Institut f\"ur Radioastronomie, Auf dem H{\"u}gel 69, 53121 Bonn, Germany}

\author[0000-0000-0000-0000]{Luigi C. Gallo}
\affiliation{Department of Astronomy and Physics, Saint Mary’s University, Halifax, NS B3H 3C3, Canada}

\author[0000-0002-2322-5232]{Jun Yang}
\affiliation{Department of Space, Earth and Environment, Chalmers University of Technology, Onsala Space Observatory, SE-439\,92 Onsala, Sweden}

\author[0000-0001-6947-5846]{Luis C. Ho}
\affiliation{Kavli Institute for Astronomy and Astrophysics, Peking University, Beijing 100871, China}
\affiliation{Department of Astronomy, School of Physics, Peking University, Beijing 100871, China}

\author[0000-0002-4455-6946]{Minfeng Gu}
\affiliation{Key Laboratory for Research in Galaxies and Cosmology, Shanghai Astronomical Observatory, Chinese Academy of Sciences, 200030 Shanghai, P.R. China}

\author[0000-0003-3389-6838]{Willem A. Baan}
\affiliation{Xinjiang Astronomical Observatory, Key Laboratory of Radio Astronomy, Chinese Academy of Sciences, 150 Science 1-Street, 830011 Urumqi, P.R. China}
\affiliation{Netherlands Institute for Radio Astronomy ASTRON, NL-7991 PD Dwingeloo, the Netherlands}

\author[0000-0003-2931-0742]{Jiri Svoboda}
\affiliation{Astronomical Institute, Academy of Sciences, Bo\^cn\'i II 1401, 14131 Prague, Czech Republic}

\author[0000-0003-4956-5742]{Ran Wang}
\affiliation{Kavli Institute for Astronomy and Astrophysics, Peking University, Beijing 100871, China}
\affiliation{Department of Astronomy, School of Physics, Peking University, Beijing 100871, China}

\author[0000-0001-9815-2579]{Xiang Liu}
\affiliation{Xinjiang Astronomical Observatory, Key Laboratory of Radio Astronomy, Chinese Academy of Sciences, 150 Science 1-Street, 830011 Urumqi, P.R. China}

\author[0000-0002-1992-5260]{Xiaoyu Hong}
\affiliation{Shanghai Astronomical Observatory, Chinese Academy of Sciences, Shanghai 200030, China}
\affiliation{Shanghai Key Laboratory of Space Navigation and Positioning Techniques, Shanghai 200030, China}
\affiliation{University of Chinese Academy of Sciences, 19A Yuquanlu, Beijing 100049, China}

\author[0000-0003-4956-5742]{Xue-Bing Wu}
\affiliation{Kavli Institute for Astronomy and Astrophysics, Peking University, Beijing 100871, China}
\affiliation{Department of Astronomy, School of Physics, Peking University, Beijing 100871, China}

\author[0000-0003-4478-2887]{Wei Zhao}
\affiliation{Shanghai Astronomical Observatory, Chinese Academy of Sciences, Shanghai 200030, China}



\begin{abstract}
Accretion of black holes at near-Eddington or super-Eddington rates is the most powerful episode that drives black hole growth, and it may work in several types of objects. However, the physics of accretion and jet-disc coupling in such a state remains unclear, mainly because the associated jets are not easily detectable due to the extremely weak emission or possibly episodic nature of the jets. Only a few near/super-Eddington systems have demonstrated radio activity, and it remains unclear whether there is a jet and what are their properties, in super-Eddington active galactic nuclei (AGNs) (and ultraluminous X-ray sources). The deficit is mainly due to the complex radio mixing between the origins of jets and others, such as star formation activity, photo-ionized gas, accretion disk wind, and coronal activity. In this work, we conducted high-resolution very long baseline interferometry (VLBI) observations to explore the jets in the highly accreting narrow-line Seyfert I system \obj. Our observations successfully revealed small-scale jets (with a linear size of $\sim45$\,parsec) at both 1.5 and 5\,GHz, based on the high radio brightness temperature, radio morphology, and spectral index distribution. Interestingly, the lack of a flat-spectrum radio core and knotty jet structures imply episodic ejections in \obj, which resemble the ejection process in Galactic X-ray binaries that are in the canonical very high state. The high accretion rates and jet properties in the AGN \obj\ may support the AGN/XRB analogy in the extreme state.
\end{abstract}



\section{Introduction\label{sec:intro}}

The Eddington ratio\footnote{$\lambda_\mathrm{Edd} \equiv L_\mathrm{bol}/L_\mathrm{Edd}$; $L_\mathrm{bol}$ is the bolometric luminosity,  $L_\mathrm{Edd}=3.2\times10^4(M_\mathrm{BH}/M_\mathrm{\odot})L_\mathrm{\odot}$ is the Eddington luminosity, and $M_\mathrm{BH}$ is the black hole mass.} is a key indicator of black hole accretion and ejection states, in both stellar-mass black holes (SBHs) and supermassive black holes (SMBHs) \citep[e.g.][]{2004MNRAS.355.1105F, 2004A&A...414..895F, 2006MNRAS.372.1366K}, which is generally $<1$ under the assumption of spherical accretion. The accretion flows and associated ejection processes with low and moderate Eddington ratios generally can be described with Advection Dominated Accretion Flows \citep[ADAFs,][]{1994ApJ...428L..13N, 1997ApJ...489..865E} and standard accretion disk \citep[or Shakura–Sunyaev disk, SSD,][]{1973A&A....24..337S}, respectively, and corresponding revisions \citep[e.g.][]{2004MNRAS.355.1105F, 2007A&ARv..15....1D, 2014ARA&A..52..529Y}.

However, super-critical accretion (with the super-Eddington accretion rate for which the black hole radiates above the Eddington luminosity) is viable in both observations and physics and potentially plays an essential role in feeding the black hole growth in the early Universe \citep[see][and references therein]{2020ApJ...904..200Y}. Furthermore, super-Eddington accretion of the first-generation SMBHs may have a deep impact on regulating the (host) galaxy evolution and the epoch of reionization through feedback processes. As accretion increases to near or super-Eddington rates, the standard disc geometry cannot be maintained and the accretion flow will inevitably evolve into a `slim disc' \citep{2013MNRAS.436...71V}. The corresponding state is sometimes called the `ultraluminous state' \citep{2009MNRAS.397.1836G}. 

Regardless of the importance and the viability of super-Eddington accretion, our understanding of the accretion and ejection processes in this accretion state remains limited, which is primarily due to that only a few systems can temporarily trigger super-Eddington accretion \citep[e.g.][]{2001Natur.414..522G, 2018ApJ...859L..20D}, and even fewer systems can maintain long-lived super-Eddington accretion \citep[e.g.][]{2021MNRAS.506.1045M, 2023NewAR..9601672K}. Even worse, the mechanism for sustaining super-Eddington accretion in those sources which persistently accrete at super-Eddington rates is unclear.

It is also widely accepted that supermassive and stellar-mass black holes have similarities in accretion physics, i.e., Active galactic nuclei (AGNs) and XRBs have similar accretion state transitions and associated ejection processes. However, it is still unclear whether the AGN/XRB analogy holds in the `ultraluminous state' and whether the geometry of the disc–corona system and jet-disc coupling are similar. Here, our interest is the connection between the short-lived canonical `very high state' (universally found in XRBs) and the long-standing super-Eddington accretion in, for example, the microquasar SS\,433, and ULXs, and which parameters are driving the long-lived super-Eddington accretion. As the time scale of state transition is proportional to black hole mass \citep{2017A&A...603A.127S, 2020ApJ...904..200Y}, a `very high state' in SMBHs (e.g. $M_\mathrm{BH}=10^7\,M_\odot$), would last $10^6$ times longer than in $10\,M_\odot$ stellar-mass black holes found in XRBs. Therefore, the study of near/super-Eddington AGNs provides an opportunity to understand the ejection process in a quasi-steady `very high state' and may shed light on the physics to sustain such a near/super-Eddington accretion. For this reason, we present very long baseline interferometry (VLBI) observations of an AGN, \obj, which radiates close to or above the Eddington limit.

\obj\ is one of the closest quasars located at a redshift of $z=0.0589$ \citep{2009ApJS..184..398H} and is regarded as an archetypal narrow-line Seyfert 1 galaxy (NLS1) based on its optical properties \citep{1983ApJ...269..352S, 2000NewAR..44..381P}. The black hole mass of \obj\ was estimated to be $M_\mathrm{BH} =9.3\times 10^6\,M_\odot$ from optical reverberation mapping \citep{2019ApJ...876..102H}. The bolometric luminosity estimated from the spectral fitting is $\log{L_{\rm bol}}=45.50-45.68$\,erg\,s$^{-1}$ \citep{2017MNRAS.468....2M}, which exceeds its Eddington luminosity with an Eddington ratio of $\lambda_\mathrm{Edd}=2.77-4.20$. Another work obtained a higher black hole mass of $M_\mathrm{BH} =2.8\times 10^7\,M_\odot$ using X-ray reverberation \citep{2021Natur.595..657W} and estimated the Eddington ratio of \obj\ as unity (or 0.3) based on the optical monochromatic luminosity (or the X-ray luminosity). However, the authors note that the luminosity could be underestimated due to photons being trapped in the disk. If we take the bolometric luminosity of $\log{L_{\rm bol}}=45.50-45.68$\,erg\,s$^{-1}$ \citep{2017MNRAS.468....2M}, which is thought to be more accurate than the estimation from the single-band luminosity, and use the larger black hole mass measurement of $M_\mathrm{BH} =2.8\times 10^7\,M_\odot$ \citep{2021Natur.595..657W}, then the Eddington ratio would be $0.92 - 1.40$. We should bear in mind that when the bolometric luminosity approaches and exceeds the Eddington luminosity, the actual mass accretion rate would be significantly higher than expected from the observable luminosity assuming a typical radiative efficiency \citep[$\eta<1$,][]{2003PASJ...55..599B} because of the photon trapping effect \citep{2000PASJ...52..499M}. The radiative efficiency of \obj\ was estimated, as one of the Palomar–Green (PG) quasars PG~0050+124, to be $\log{\eta}=-2.21$ or $-1.18 \pm 0.04$ \citep[based on the mass estimates with the broad emission line widths and the $M-\sigma_*$ correlation, respectively, see][]{2011ApJ...728...98D}. With such a high Eddington ratio and low radiative efficiency, therefore, the SMBH in \obj\ must be growing with a mass accretion rate notably higher than the Eddington limit, i.e. the super-Eddington accretion rate.

\obj\ is also an extremely radio-quiet AGN with radio loudness parameter $\mathcal{R}=0.35$ (fn.\footnote{$\mathcal{R}=1.3\times10^5(L_{5}/L_\mathrm{B})$; $L_5$ is the radio luminosity at 5\,GHz and $L_\mathrm{B}$ is the optical luminosity of the nucleus at $\lambda_B = 4400$\,\AA; radio-quiet AGNs are defined as $\mathcal{R}<10$.}) \citep{2020ApJ...904..200Y}. Radio emission from radio-quiet AGNs is complex and remains a subject of debate \citep{2019NatAs...3..387P}. On the other hand, the presence of jets in near or super-Eddington systems and how it is launched are also questions that need to be explored. X-ray observations of \obj\ indicate that the X-ray corona exhibits some structure and part of it may be collimated and ejected \citep{2007MNRAS.377.1375G, 2017MNRAS.471.4436W}. \obj\ is an extreme example of a nearby highly accreting and radio-quiet AGN, providing an ideal laboratory for studying outflow activities\footnote{throughout the work, we use `outflow' to refer to both collimated jets and wind-like outflows.} including jets and winds, with high spatial resolutions.

In this work, we report the Very Long Baseline Array (VLBA) and European VLBI Network (EVN) plus the enhanced Multi-Element Remote-Linked Interferometer Network ($e$-MERLIN) observations of the nuclear region in \obj\ and we also analyze the archival data from VLA and MERLIN. Our paper is organized as follows, Section \ref{sec:obs} details the multi-band observations, data reduction, and analysis of the target \obj, while Section \ref{sec:dis} presents the results and discussions. Finally, we provide our conclusions in Section \ref{sec:conclusion}. Throughout this work, we adopt the standard $\Lambda$CDM cosmology with $H_0 =71$ km s$^{-1}$ Mpc$^{-1}$, $\Omega_\Lambda=0.73$, $\Omega_m=0.27$, and the corresponding physical scale is 1.125\,pc\,mas$^{-1}$ in \obj.

\section{Observations and data reduction} \label{sec:obs}

\subsection{VLBI observation and data reduction} \label{vlbiobs}
We observed \obj\ on 2018 September 23 with 10 antennas of the Very Long Baseline Array (VLBA) and on 2020 November 17 with 19 antennas of the European VLBI Network (EVN) plus the enhanced Multi-Element Remote-Linked Interferometer Network ($e$-MERLIN). The VLBA observation was carried out at L-band ($1.548$\,GHz or 1.5\,GHz for short, the project code BY145), and the EVN+$e$-MERLIN observation was conducted at C-band ($4.926$\,GHz or 5\,GHz for short, the project code EY037), respectively. The total VLBA observing time is 2 h with a data recording rate of 2\,Gbps, and the total time of the EVN+$e$-MERLIN observation is 8 h with a data recording rate of 4\,Gbps. Both observations were performed in the phase-referencing mode, using J0056$+$1341 (R.A.: 00$^\mathrm{h}$56$^\mathrm{m}$14.816010$^\mathrm{s}\pm0.000013^\mathrm{s}$, Dec.: $+$13$^\circ$41$^\prime$15.75506$^{\prime\prime}\pm$0.00044$^{\prime\prime}$) as the phase reference calibrator. 

We calibrated the VLBI data in the Astronomical Image Processing System (\softw{aips}), a software package developed by the National Radio Astronomy Observatory (NRAO)  of U.S. \citep{2003ASSL..285..109G}, following the standard procedure. A-prior amplitude calibration was performed using the system temperatures and the antenna gain curves provided by each station. The earth orientation parameters were obtained and corrected using the measurements from the U.S. Naval Observatory database and the ionospheric dispersive delays were corrected based on a map of the total electron content provided by the GPS satellite observations \footnote{\url{https://cddis.nasa.gov}}. The opacity and parallactic angles were also corrected based on the auxiliary files attached to the data. The delay of the visibility phase and the telescope bandpass were calibrated using the bright radio source 3C~454.3. Next, we performed a global fringe-fitting on the phase-referencing calibrator, J0056$+$1341, by assuming a point source model to solve miscellaneous phase delays. 

The phase calibrator J0056$+$1341 shows a core-jet structure that extends upto $\sim$100 mas to the north  (see supplementary Figure \ref{fig:j0056} in Appendix \ref{apd:tab} for their $1.5$ and $5$\,GHz images). We performed self-calibration to the phase calibrator and obtained its CLEAN model, which was then used as the input model to re-solve the phases in \softw{aips}. This operation can eliminate phase reference errors due to the jet structure. Finally, both phase and amplitude solutions obtained from the phase calibrator were applied to the target \obj. The calibrated uv-data was exported to \softw{difmap} \citep{1997ASPC..125...77S} for deconvolution. Based on a signal-to-noise ratio of $\sim30$ and $\sim14$ in $1.5$ and $5$\,GHz in the residual map, respectively, we decide to not perform self-calibration to the target source.

\begin{figure*}
\centering
\includegraphics[scale=0.4]{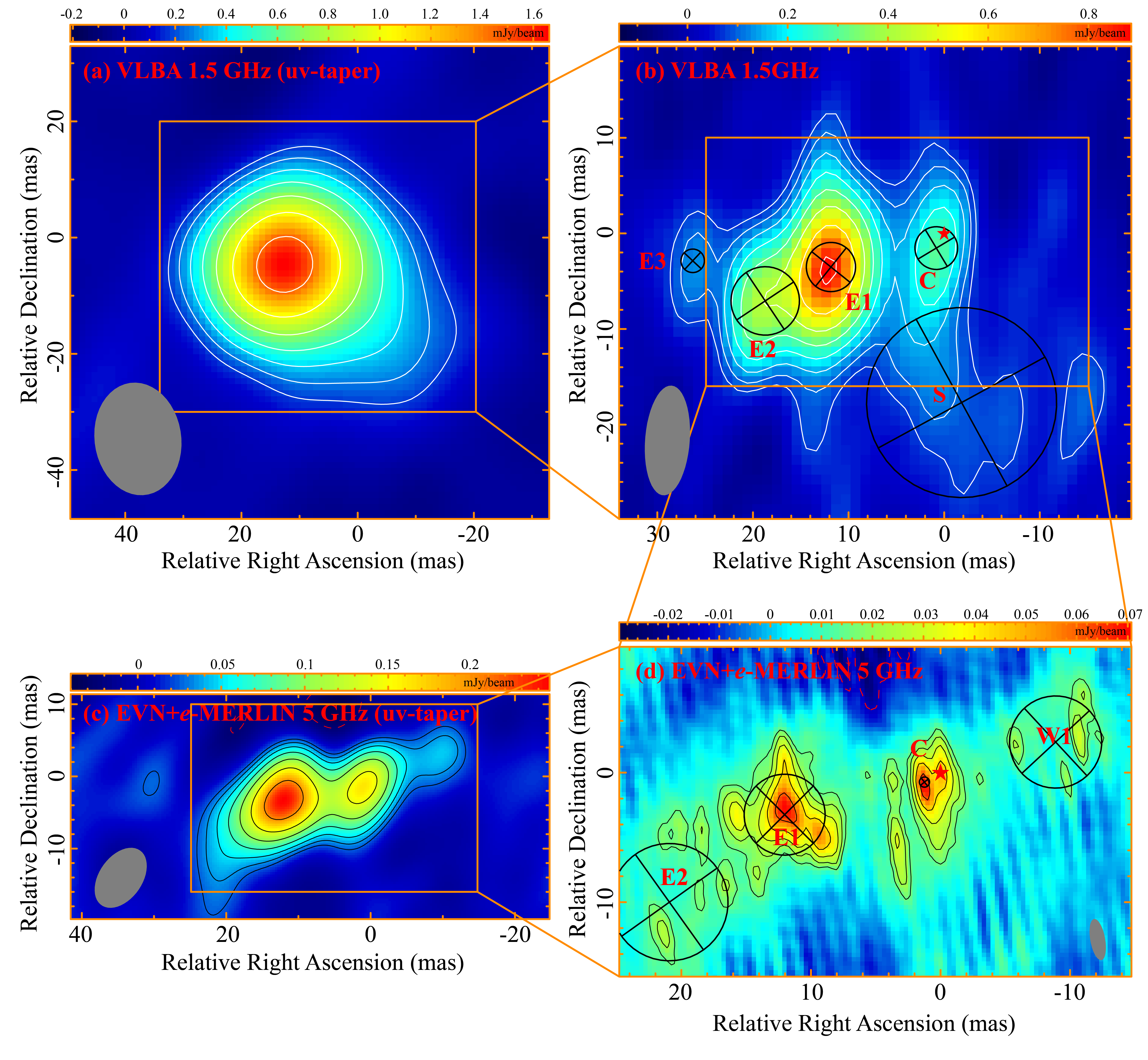}
\caption{\textbf{1.5 and 5\,GHz VLBI images of \obj}. All the images are produced with natural weight and the map reference is at the \telescope{Gaia} position. The restoring beams are displayed as grey ellipses in the lower-left/right corner of each panel, which are $19.9\times14.9$, $11.5\times4.67$, $9.27\times5.95$ and $3.22\times1.14$\,mas from panel (a) to (d), respectively. The contours are at 3$\sigma\times (-1, 1, 1.41, 2, 2.83,\dots)$, where positive contours are white and black solid and negative ones are red dashed. From panel (a) to (d), the image peak is $1.66$, $0.88$, $0.24$, and $0.10$\,mJy/beam, respectively, and the rms noise is $0.06$, $0.025$, $0.008$ and $0.007$\,mJy/beam, respectively. The circles with crosses inside indicate the corresponding Gaussian models. In panels (b) and (c), the red stars mark the \telescope{Gaia} optical positions. The uncertainty in the \telescope{Gaia} position is $\Delta\alpha=0.18$ and $\Delta\delta=0.17$\,mas, which includes an astrometric excess noise error of $0.14\,\mathrm{mas}$. At the redshift of \obj, 1\,mas corresponding to 1.125\,pc.}\label{fig:vlbi}
\end{figure*}

We performed different deconvolution algorithms in \softw{difmap} to produce radio maps, i.e. CLEAN and Gaussian model-fit. It is important to note that the solutions for visibilities are not necessarily unique when complex structures are to be handled. In order to compare the goodness of the deconvolution results, we summarised statistical parameters in Table \ref{tab:sta}. Obviously, the statistical parameters of the Gaussian model-fit are close to the CLEAN, while the CLEAN is better than the Gaussian model-fit based on the $\chi^2_r$-value (especially at 5\,GHz), which is reasonable as the emission regions are generally more complex than the representation of a few Gaussian components. The resulting CLEAN images and Gaussian model-fitted components are shown in Figure \ref{fig:vlbi}. The root-mean-square (rms) noise in the residual map (Column 3 of Table \ref{tab:sta}) is larger than the off-source rms noise in the map (Column 4 of Table \ref{tab:sta}), indicating that some diffuse emission cannot be recovered in the full resolution map. The off-source rms represent the background noise fluctuation, whereas the rms of the residual map incorporate residual flux densities. Therefore, the loss of flux density in the images can be characterized as $\frac{\sigma_r-\sigma_{rms}}{\sigma_{rms}}$ (Column 5 of Table \ref{tab:sta}). Simultaneously, we also produced a uv-tapered map to reduce the weight of the long-baseline visibilities (and thus also reduce the resolution) in an attempt to recover the weak and extended emission, see panels (a) and (c) in Figure \ref{fig:vlbi}.

%\subsection{Archived MERLIN data} \label{sec:merlin}
%Two MERLIN C-band ($4.99$\,GHz) datasets of the target \obj\ are available in the MERLIN data archive \footnote{\url{http://www.merlin.ac.uk/archive/acknowledge.html}\\\url{emerlin.support@jb.man.ac.uk}}. These data were observed on 1996 December 1 with six antennas: Defford (De), Cambridge (Ca), Knockin (Kn), Darnhall (Da), Mark II (Mk), and Tabley (Ta), and 1997 November 6 with five antennas (De, Ca, Kn, Da, Ta). The two MERLIN observations provide a minimum baseline of 0.07 M$\lambda$ ($\sim$4\,km) and 0.47 M$\lambda$ ($\sim$28\,km), respectively.  Both observations have a maximum baseline of 3.62 M$\lambda$ ($\sim$217\,km) with a resolution of $\sim$0.1\,arcsec. The calibrator J0106$+$1300 (RA: 01$^\mathrm{h}$06$^\mathrm{m}$33.3558$^\mathrm{s}$, Dec: $+$13$^\circ$00$^\prime$02.608$^{\prime\prime}$) was used for phase reference, and 3C~286 and OQ~208 were used as the primary flux density scale and bandpass calibrator for both datasets. The archived MERLIN data has already been calibrated in AIPS, including preliminary bandpass and flux density scale calibration, as well as phase and amplitude calibrations. We imported the calibrated data into DIFMAP for manual imaging and self-calibration. In both datasets, the target is detected with a signal-to-noise ratio above $7$, but self-calibration was not performed. A two-dimensional Gaussian model was fitted to the visibility data to obtain the integrated and peak flux density, which are shown in Table \ref{tab:archive}. 

\subsection{Archived VLA data} \label{sec:vla}
We retrieved the raw visibility data of \obj\ observed by the Very Large Array (VLA) from the NRAO data archive \footnote{\url{https://archive.nrao.edu/archive/advquery.jsp}}, including historical VLA and the newly observed Karl G. Jansky VLA (JVLA) data. Although some data have been published (see Table \ref{tab:archive}), to ensure consistency in the data reduction, we performed a manual calibration for all available datasets using the Common Astronomy Software Application \citep[CASA v5.1.1,][]{2007ASPC..376..127M}. Our data reduction followed the standard routines described in the CASA cookbook. We adopted the `Perley-Butler 2017' flux density standard to set the overall flux density scale for the primary flux calibrator and then bootstrapped the secondary flux density calibrators and the target. For the historical VLA datasets, we determined the gain solutions using a nearby phase calibrator and transferred them to the target \obj. For the JVLA datasets, we also determined antenna delay and bandpass by fringe-fitting the visibilities. For the data observed after 1998, we performed an ionosphere correction using the data obtained from the CDDIS archive. Deconvolution, self-calibration, and model-fit were performed in \softw{difmap}. The final images were created using natural weight. Due to the good uv-coverage, simple emission structure, and high signal-to-noise ratio (SNR$>$9), the VLA data allows for self-calibration using a well-established model. For data with lower SNR, we used three times the image noise as the upper limit for the flux density.

\subsection{Astrometry of the VLBI data} \label{sec:astrometry}

We measure the uncertainties of the astrometric measurements from three main origins: (1) Positional uncertainties of phase-referencing calibrators. In phase-referencing observations, the coordinates of the target are referenced to a close calibrator. The calibrator J0056$+$1341 was selected from the catalog rfc\_2022a in Astrogeo \footnote{\url{http://astrogeo.org/}} with precise position with accuracies $\Delta\alpha=0.20$\,mas in right ascension and $\Delta\delta=0.44$\,mas in declination; (2) Astrometric accuracy of phase-referenced observations. Primarily concerning the station coordinate, Earth orientation, and troposphere parameter uncertainties, which can be measured through the formula and data from \citet{2006A&A...452.1099P}. This portion contribute position errors $\Delta\alpha\approx0.26$\,mas and $\Delta\delta\approx0.50$\,mas in VLBA 1.5\,GHz observation and $\Delta\alpha\approx0.27$\,mas and $\Delta\delta\approx0.47$\,mas in EVN+$e$-MERLIN 5\,GHz observation; (3) Thermal error due to the random noise \citep[e.g.][]{1986isra.book.....T, 2017AJ....153..105R}. This uncertainty can be characterized as $\sigma_t\sim\theta_\mathrm{B}/(2\times$SNR$)$, where $\theta_\mathrm{B}$ is the full-width at half maximum of the restoring beam and SNR is the signal-to-noise ratio. In this work, we take this value from \softw{difmap}.

During the self-calibration process, the absolute coordinate position of the phase-referencing calibrator is lost and the brightest feature of the image is shifted to the phase center of the map. In general, due to the frequency-dependent shift in the peak of the optically thick component and the slightly different distribution of the radio emission at different resolutions, the brightest component may not be the same component from one observation to the next. This would induce a systematic offset between two images. The alignment between the images of two frequency observations can be done using an optically thin component as a reference since its position is less affected by the frequency-dependent opacity effect \citep[e.g.][]{2001ApJ...550..160M, 2008A&A...483..759K,2011A&A...532A..38S,2013A&A...557A.105F}.

In our VLBI observations, we used J0056$+$1341 as the phase-referencing calibrator, which has a flat-spectrum radio core. We obtained VLBI C (4.34\,GHz) and X-band (7.62\,GHz) data from Astrogeo. As neither the C and X-band data are self-calibrated, the core shift at C-band can be directly estimated as $\Delta$R.A. $\sim0.69\,\mathrm{mas}$, $\Delta$DEC. $\sim-0.29\,\mathrm{mas}$ relative to the X-band data (see supplementary Table \ref{tab:coord} in Appendix \ref{apd:tab}). J0056$+$1341 also has a significant offset between C and L bands (see supplementary Figure \ref{fig:j0056} in Appendix \ref{apd:tab}) in our observations, and the offset of the VLBA L-band image is determined using the optically thin component J1 of the jet, estimated as $\Delta$R.A. $=1.742\,\mathrm{mas}$ and $\Delta$DEC. $=-3.228\,\mathrm{mas}$ (see supplementary Table \ref{tab:coord} in Appendix \ref{apd:tab}) relative to the EVN+$e$-MERLIN C-band image.

For the target \obj, we use the position of the brightest optically thin ($\alpha=-0.86\pm0.07$) component in the tapered EVN+$e$-MERLIN 5\,GHz image (panel c of Figure \ref{fig:vlbi}) to align with the VLBA 1.5\,GHz image. The peak position of the 1.5\,GHz image was moved in \softw{difmap} to align with the 5\,GHz image by $\Delta$R.A. $=1.33\pm0.38\,\mathrm{mas}$, $\Delta$DEC. $=-0.77\pm0.69\,\mathrm{mas}$, where the positional uncertainties accounts for both 1.5 and 5\,GHz astrometric uncertainties of the brightest component E1. The centroid of the optical emission obtained from the second data release \footnote{\url{https://gea.esac.esa.int/archive/}} of the \telescope{Gaia} mission \citep{2018A&A...616A..11G, 2018A&A...616A...1G} is R.A.$=00^\mathrm{h}53^\mathrm{m}34^\mathrm{s}.933288\pm0.000012$ and DEC.$=+12^\mathrm{\circ}41^\prime35^{\prime\prime}.93081\pm0.00017$ (J2000). This includes astrometric excess noise error of $0.14\,\mathrm{mas}$. Coordinates for the target are also listed in supplementary Table \ref{tab:coord} in Appendix \ref{apd:tab}.

\begin{figure}
\centering
\includegraphics[scale=0.45]{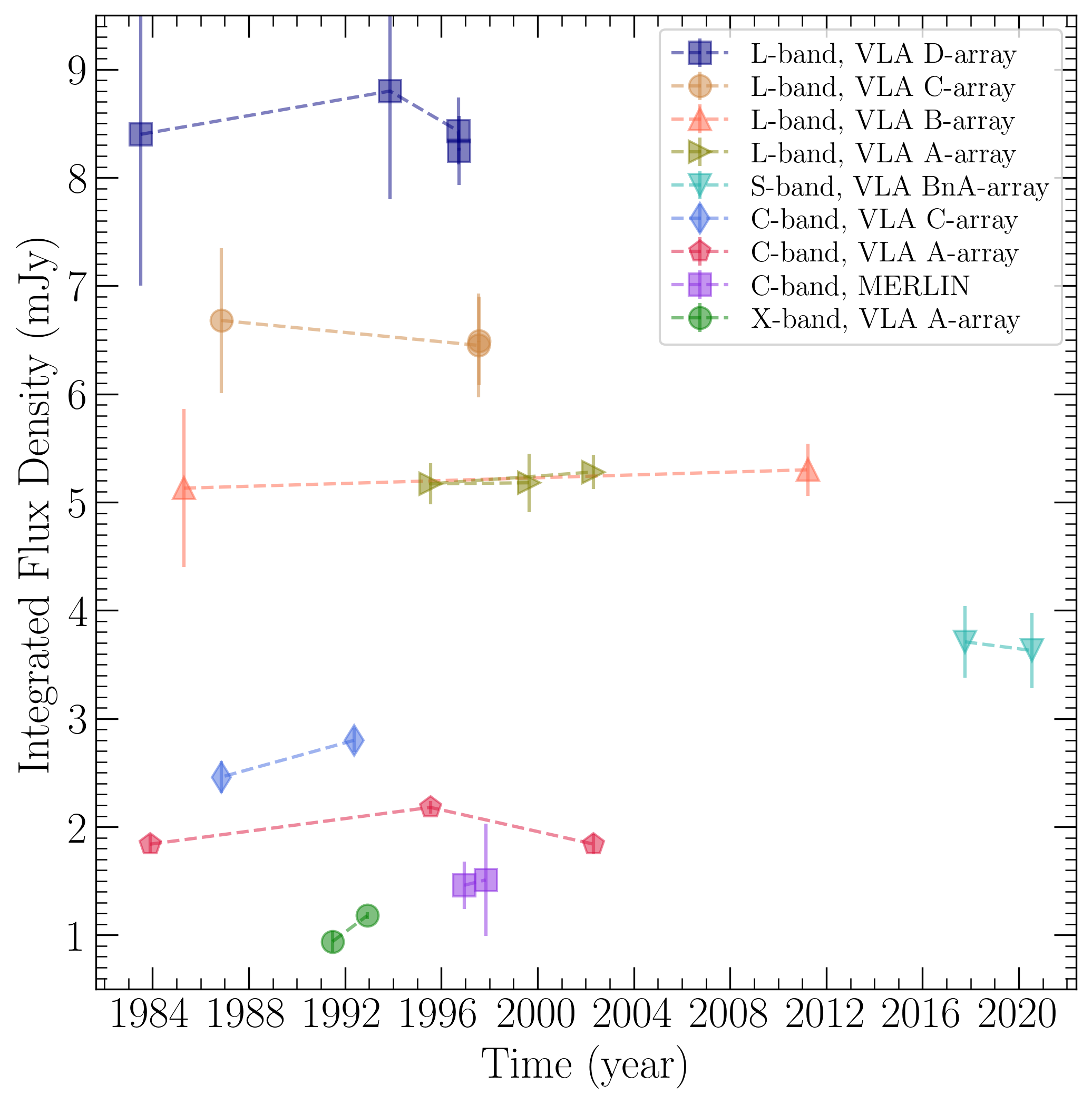}
\caption{The radio light curves of \obj\ over a time interval of 37 years. The integrated radio flux densities and their uncertainties are taken from Table \ref{tab:archive}, where the data with the same observing band (approximately equal central frequencies) and arrays/sub-arrays are concatenated to show the variability.}\label{fig:var}
\end{figure}

\begin{figure}
\centering
\includegraphics[scale=0.41]{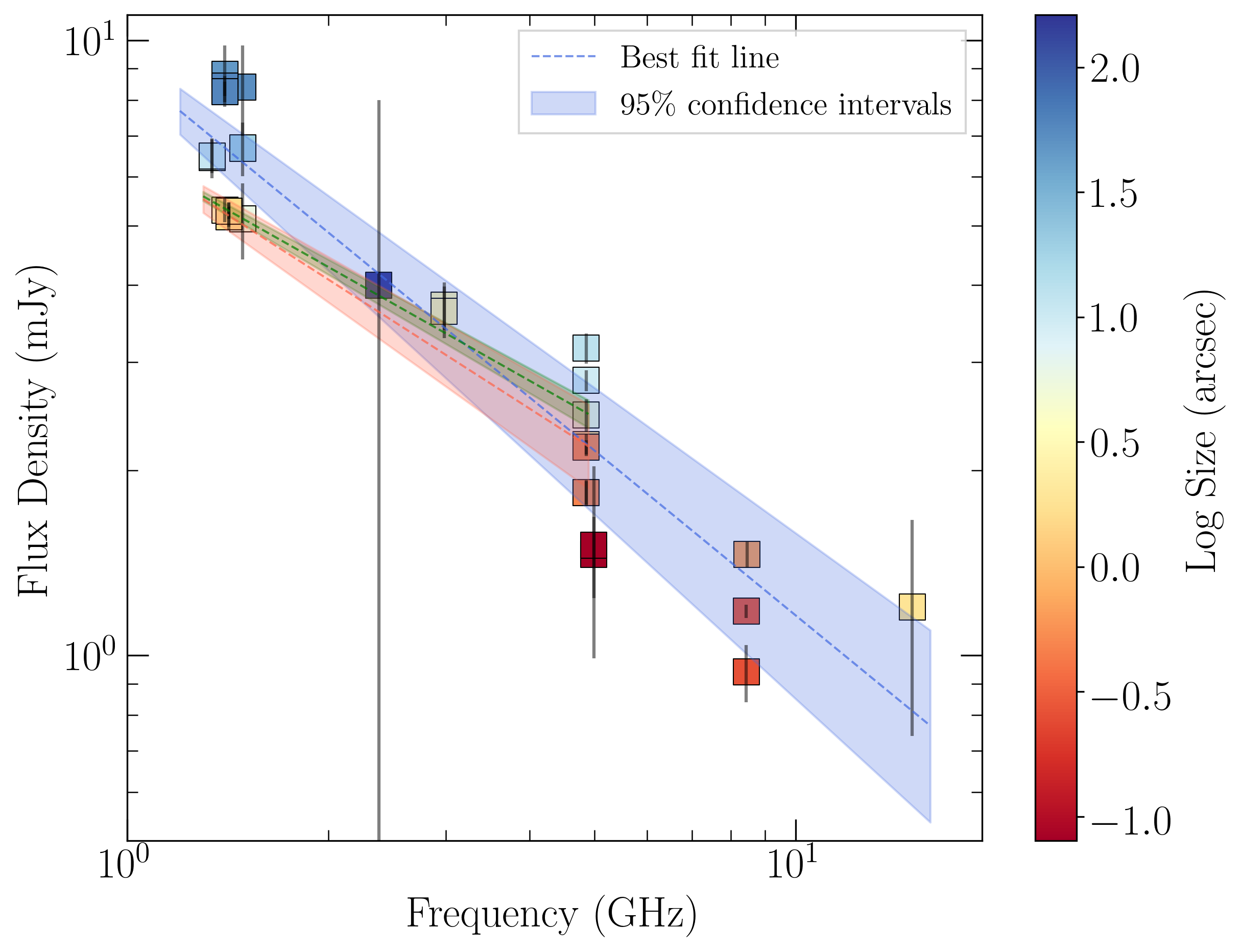}
\caption{Wide-band radio spectrum of \obj. The integrated radio flux density measurements of \obj\ in five radio frequency bands between 1.4 and 15\,GHz are shown, where the flux density and uncertainties are taken from Table \ref{tab:archive}. The blue dashed line is the model-fitting result with a power-law spectrum using all the data points presented here. The power-law slope (spectral index) is $-0.89$ and the blue belt shows the 95\% confidence interval ($0.10$). The green and red dashed lines show the power-law fitting between 1.4 and 5\,GHz datasets with similar size scales, i.e. 1.3$\sim$1.5 (arcsec, red) and 4.3$\sim$5.3 (arcsec, green), respectively. The green and red belts indicate their 95\% confidence intervals.}\label{fig:si}
\end{figure}

\begin{figure*}
\centering
\includegraphics[scale=0.55]{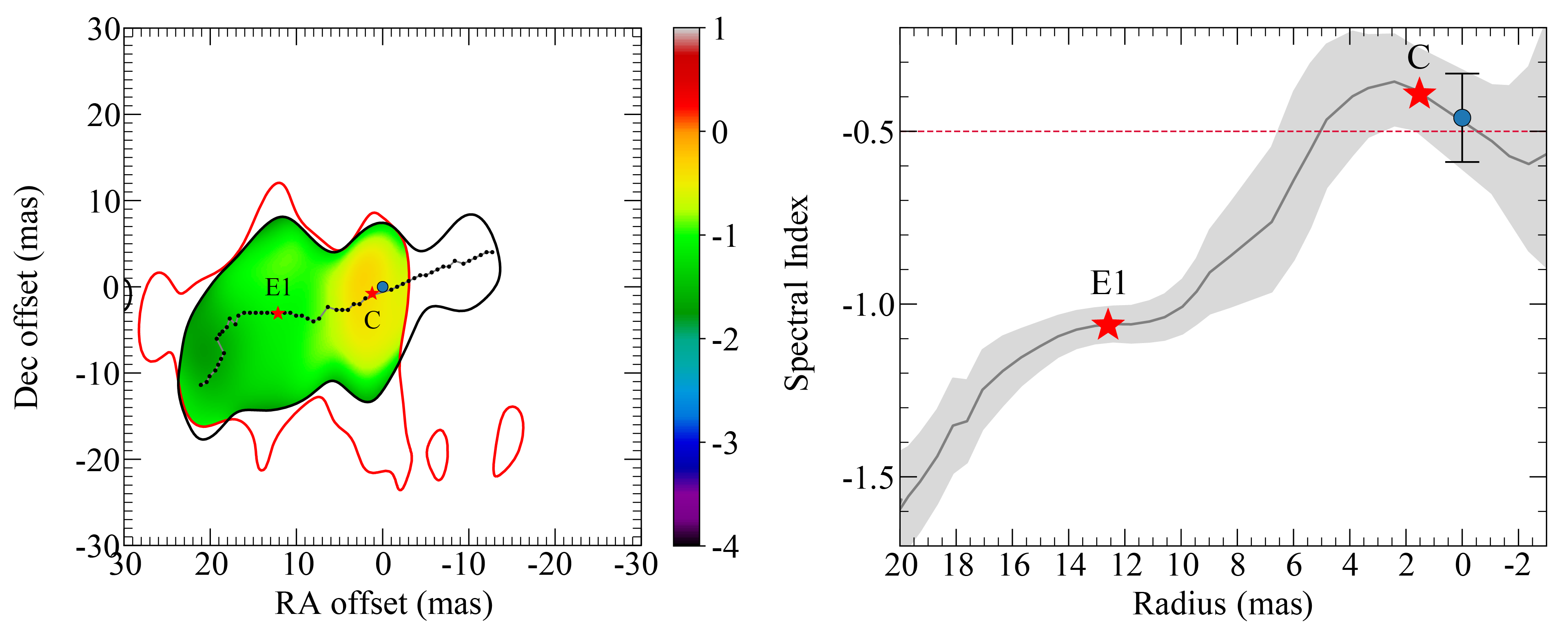}
\caption{1.5--5\,GHz spectral index distribution of \obj\ on the parsec scale. Left panel: the spectral index map produced by using the naturally weighted clean map at 1.5 and 5\,GHz. The region with radio flux density below 3$\sigma$ was set as blank (white), i.e. the outer region of the red (for 1.5\,GHz) and black (for 5\,GHz) curves. Radio spectral indices within both red and black curves are reliable. The black dots and the grey line indicate the ridgeline obtained from the tapered 5\,GHz EVN+$e$-MERLIN image. The red stars indicate the centroid positions of the Gaussian components E1 and C from the 5\,GHz image. Right panel: the spectral index distribution along the ridgeline. A positive radius corresponds to positive right ascension coordinates, and vice versa. The \telescope{Gaia} position is set as the reference. The grey belt marks the uncertainty of the spectral indexes along the ridgeline. The red stars mark the locations of the Gaussian components E1 and C from the 5\,GHz image. In both panels the blue asterisks indicate the \telescope{Gaia} position.}\label{fig:spmap}
\end{figure*}

\subsection{Radio Spectrum} \label{sec:spectrum}
To obtain the radio spectral index, we first checked the variability of \obj. Figure \ref{fig:var} shows the radio flux density versus the observing epoch. The largest variability we identify is $\sim$8\%, which is from the VLA A-array observations at the C-band between epochs 1983 and 1995. Since there is no evidence for extreme variability on a time scales of $\sim$30\,years, we plot the radio flux density versus the frequency in Figure \ref{fig:si} using all archival data. The least-square fitting gives an overall radio spectral index of $-0.89\pm0.10$. From Figure \ref{fig:si}, we can see the radio flux density changes with the collection area. The radio spectral index between 1.4 and 5\,GHz, using the datasets with a similar resolution (\textit{i.e.}, $1.3\sim1.5$ and $4.3\sim5.3$\,arcsec), is $-0.69\pm0.08$ and $-0.61\pm0.02$, respectively. These yields flatter spectral indices than the overall fit. Given the total flux density in Section \ref{sec:error}, the spectral index between VLBA 1.5 and EVN+$e$-MERLIN 5\,GHz is $-1.06\pm0.13$, consistent with the overall radio spectral index.

\begin{figure}
\centering
\includegraphics[scale=0.4]{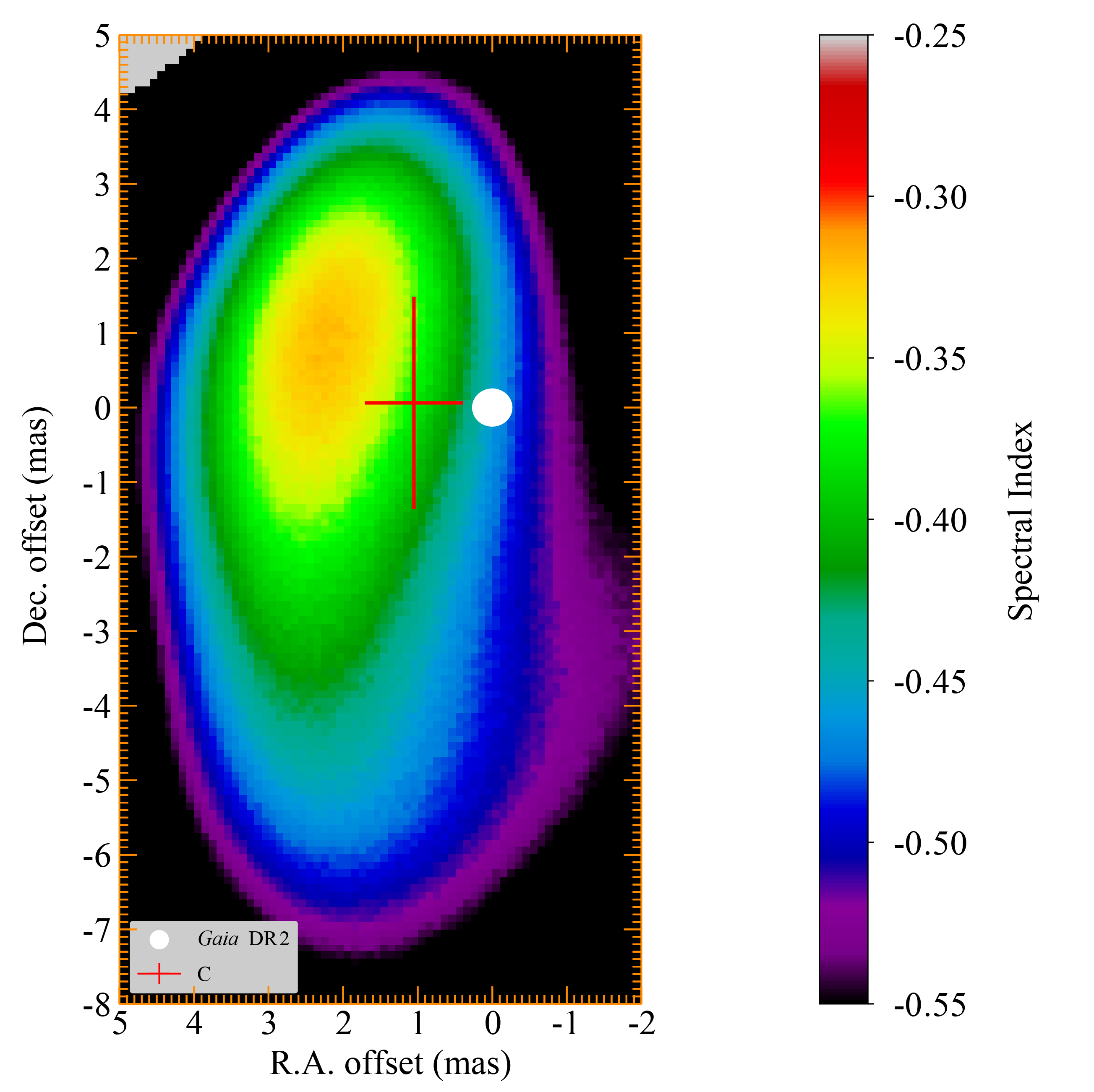}
\caption{Comparison of positions between $Gaia$ DR2 and VLBI component C in spectral index map. Here only the flat spectral region ($\alpha>-0.5$) is shown. We take $3\sigma$ position error of component C here, where $1\sigma$ position error is estimated through the method described in Section \ref{sec:astrometry}. \label{fig:astrometry}}
\end{figure}

To obtain the spectral index distribution for the high-resolution data observed with the VLBA and EVN+$e$-MERLIN, we created a spectral index map following the procedure described in \citet{2014AJ....147..143H}. Here, we used the uv-tapered image by 0.5 at 20 $\mathrm{M}\lambda$ at 5\,GHz and restored it to match the 1.5\,GHz map. Similarly, the alignment of two images is through the brightest optically thin component E1. The spectral index was calculated pixel-by-pixel between the $1.5$ and $5$\,GHz total intensity maps. For a given frequency, pixels with an intensity less than $3\sigma_{rms}$ were removed. The spectral index map between $1.5$ and $5$\,GHz is shown in the left panel of Figure \ref{fig:spmap}. Both $Gaia$ position and component C are in the flat spectral region ($\alpha>-0.5$, see also Figure \ref{fig:astrometry}.)

We study the radio spectral index distribution along the jet trajectory by estimating a ridge line of the jet. We define the jet ridge as the line that connects the peaks of one-dimensional Gaussian profiles fitted to the profile (slices) of the jet brightness drawn orthogonally to the jet direction \citep[see][]{2020A&A...641A..40V}. To obtain the ridge line, we performed a fitting to the tapered and restored (with circular beam) 5\,GHz EVN+$e$-MERLIN image. The first slice starts from the \telescope{Gaia} position and a position angle of 220$^\circ$ (regarding to the East direction anticlockwisely) was initially adopted for the jet direction. The step between individual slices was set to be smaller than the beam size (a measure taken in order to ensure the continuity of the brightness profiles recovered from adjacent slices). The jet direction was readjusted among every three ridge line points. The resulting jet ridge line and spectral index distribution are presented in Figure \ref{fig:spmap}.

\subsection{Radio flux density and Brightness Temperature} \label{sec:error}

We estimate flux density uncertainties following the instructions described by \citet{1999ASPC..180..301F}. In this work, the integrated flux densities $S_i$ were extracted from Gaussian model-fit in \softw{difmap}, where a standard deviation in model-fit was estimated for each component and considered as the fitting noise error. Additionally, we assign the standard 5 and 10\% errors originating from amplitude calibration of VLBA (see VLBA Observational Status Summary 2018B \footnote{\url{https://science.nrao.edu/facilities/vlba/docs/manuals/oss2018B}}) and EVN \citep[e.g.][]{2018A&A...619A..48R} data, respectively.

The radio brightness temperature was estimated using the formula \citep{2005ApJ...621..123U}: 
\begin{equation}\label{eq:bt}
T_\mathrm{B}=1.8\times10^9(1+z)\frac{S_i}{\nu^2\phi_{min}\phi_{maj}}~\mathrm{(K)},
\end{equation}
where \(S_i\) is the integrated flux density of each Gaussian model component in units of mJy (column 5 of Table \ref{tab:vlbi}); $\phi_{min}$ and $\phi_{maj}$ is minor- and major-axis of the Gaussian model (i.e. $\mathrm{FWHM}$) or the restored beam in milli-arcsec; \(\nu\) is the observing frequency in GHz (column 2 of Table \ref{tab:vlbi}), and \(z\) is the redshift.

The resolution limit $\theta_{lim}$ for Gaussian components can be estimated using the formula \citep{2005astro.ph..3225L},
\begin{equation}
\theta_{lim}=2^{2-\beta/2}\theta_B\left(\frac{\ln{(2)}}{\pi}\ln{\frac{\mathrm{SNR}}{\mathrm{SNR}-1}}\right)^{1/2}, 
\end{equation}
where $\theta_B$ is the FWHM of the synthesized beam, SNR is the signal-to-noise ratio and $\beta=2$ for the natural weight. If the fitted component size is lower than the corresponding resolution limit, then $\theta_{lim}$ was used instead as the component size, and the component was identified as unresolved (or the radio-emitting region cannot be constrained). The resolution limits for each component are listed in Column 7 of Table \ref{tab:vlbi} and the estimated $1.5$\,GHz and $5$\,GHz radio brightness temperatures are listed in Column 8 of Table \ref{tab:vlbi}. Since the measured component size is only the upper limit, the radio brightness temperature should be considered as the lower limit. We also estimated a total radio flux density from a uv-tapered image, which is 2.636$\pm$0.282 and 0.765$\pm$0.085\,mJy for 1.5 and 5\,GHz, respectively, and the corresponding source angular sizes are $\sim$50 and $\sim$40\,mas, respectively.

\section{Results and Discussion} \label{sec:dis}

\subsection{A canonical jet in the super-Eddington AGN \obj.}
Panel (b) of Figure \ref{fig:vlbi} shows the 1.5\,GHz VLBA image, displaying a quasi-continuous emission structure elongated along the east-west direction with an extent of $\sim$45\,parsec (pc). Panel (d) of Figure \ref{fig:vlbi} shows a higher-resolution (the beam FWHM is $3.22\times1.14$\,mas) image obtained from the 5\,GHz EVN+$e$-MERLIN observation. The bright components in the 1.5\,GHz image are resolved into a series of knots in the 5\,GHz image. Most of these components (except for components S and W1) have brightness temperatures $>10^7$\,K (Table \ref{tab:vlbi}), and the whole radio emitting structure has an overall steep spectrum (Figure \ref{fig:si}), which favors a jet origin and is unlikely from star-forming activities \citep{1991ApJ...378...65C} and thermal free-free radiation of the hot molecular disc surrounding AGNs \citep{1997Natur.388..852G}. Furthermore, based on the identification of optical and radio cores (see below), the bilateral radio structures in the 5\,GHz EVN image (panel d of Figure \ref{fig:vlbi}) are consistent with the assembling of approaching and receding jets. Correspondingly, the brightness asymmetry between the two branches of the bilateral structures indicates the Doppler boosting effect of relativistic jets.

\begin{figure*}
\centering
\includegraphics[scale=0.5]{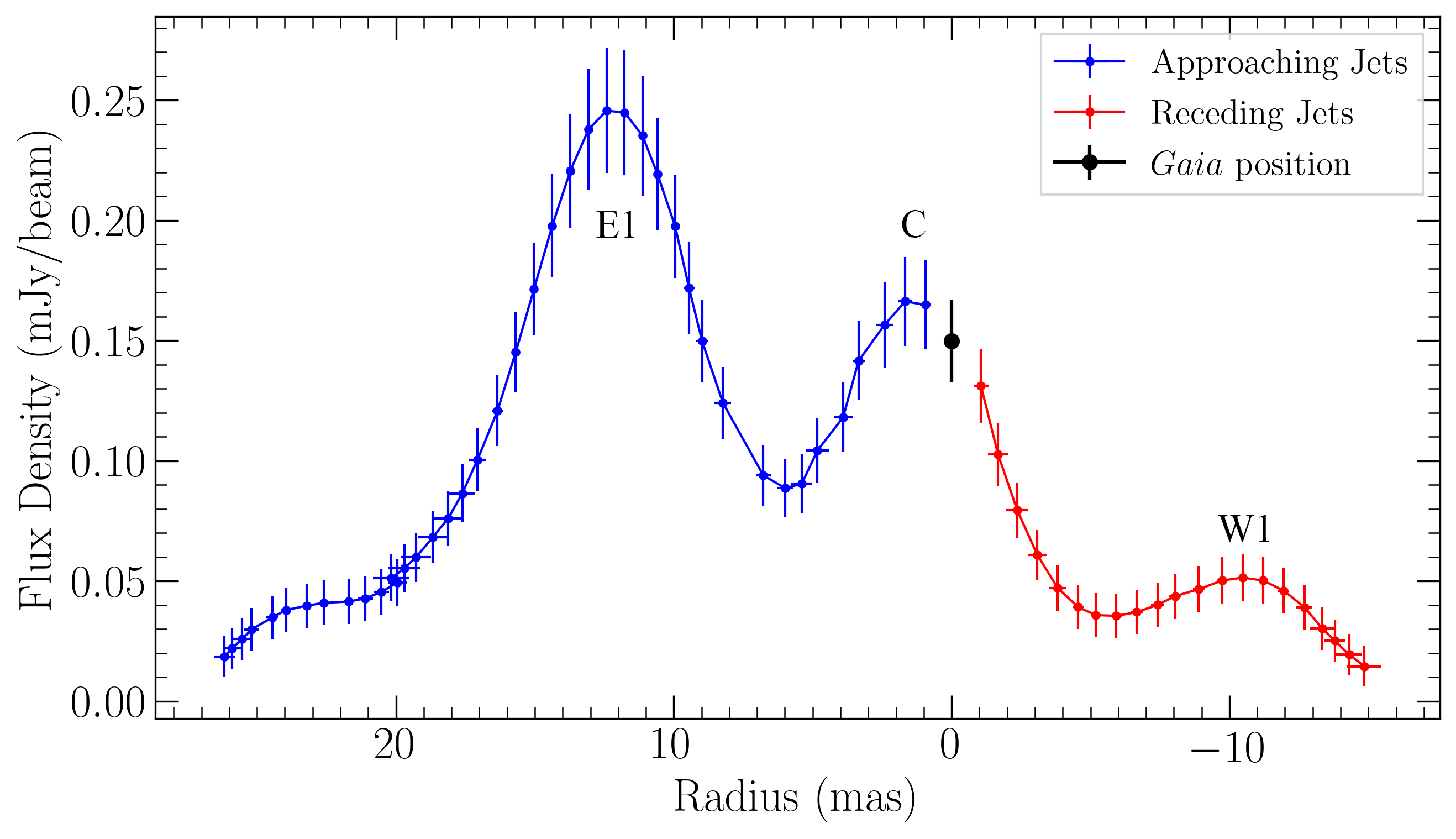}
\caption{5\,GHz flux density distribution along jet ridge line for \obj. The data points with positive (blue) and negative (red) radii are the approaching and receding jets, respectively. Positional uncertainties were directly measured in fitting the ridge line in Section \ref{sec:spectrum} and flux uncertainties are estimated by accounting for thermal noise errors and calibration uncertainties (see Section \ref{sec:error}).}\label{fig:ridgeline}
\end{figure*}

The jet base of an AGN typically has a flat radio spectrum due to the synchrotron self-absorption in the optically-thick region ($\alpha>-0.5$, fn.\footnote{$S_\nu\propto\nu^\alpha$; $S_\nu$ is the flux density and $\nu$ is the frequency}). The radio core of \obj\ is likely close to component C because it is located in a relatively flat spectral region (see Figure \ref{fig:spmap}), has a high radio brightness temperature of $\log{(T_\mathrm{B}/{\rm K})}>7$ in both 1.5 and 5\,GHz (see Table \ref{tab:vlbi}), and is roughly associated with the \telescope{Gaia} position. The 5\,GHz image further resolves component C and measures a more accurate position of its peak, which slightly offsets from the \telescope{Gaia} position (see panel d of Figure \ref{fig:vlbi} and Figure \ref{fig:astrometry}). Actually, it was already shown that there is a significant offset between VLBI and \telescope{Gaia} positions in AGNs \citep{2017MNRAS.467L..71P, 2017A&A...598L...1K, 2019ApJ...871..143P}. Interestingly, the offset between component C and the \telescope{Gaia} position is consistent with the observations of Seyfert I galaxies \citep[i.e. the upstream offsets of \telescope{Gaia} positions correspond to VLBI positions, see][]{2019ApJ...871..143P}. Therefore, the VLBI/\telescope{Gaia} offset in \obj\ can be attributed to the dominance of the accretion disk in the optical band (the \telescope{Gaia} position), while the VLBI observations alternatively track the emissions of jets \citep{2019ApJ...871..143P}.

Based on both the radio and optical cores, there is an obvious spectral steepening along the eastern jets, while it is less prominent in the western jet (see the right panel of Figure \ref{fig:spmap}). This property resembles canonical jets in blazars and can be explained with radiative losses of the synchrotron-emitting plasma or with the evolution of the high-energy cutoff in the electron spectrum \citep{2014AJ....147..143H}.

\subsection{Is the ejection process in \obj\ episodic?}
Component C is most likely the closest component to the radio core, however, the spectrum of C itself is too steep (as a reference, the spectral index at the peak position of component C is roughly $-0.4$ from the spectral index map, see Figure \ref{fig:astrometry}) to be from the nucleus of \obj. Taking the integrated flux density of C at 1.5 and 5\,GHz (see supplementary Table \ref{tab:vlbi} in Appendix \ref{apd:tab}), we can estimate the $1.5-5$\,GHz spectral index of component C as $-0.88\pm0.11$. Recalling the offset between the VLBI position of component C and the \telescope{Gaia} position of the optical nucleus, the natural explanation of component C is nascent ejecta.

Interestingly, the positions of local brightness enhancements (the knots E1 and W1) are symmetric (see Figure \ref{fig:ridgeline}) and components E1 and C seem to be embedded on a continuous background of the jet stream and mimic discrete blobs/knots. On the origin of the knots, several works envisage them as the shocks traveling along the jets \citep[e.g.][]{2014AJ....147..143H}, while the other works interpret them as discrete blobs from episodic ejection \citep[e.g.][]{2019ApJ...877..130S}. We note that the relatively flatter spectra around the Gaussian components (C and E1) in Figure \ref{fig:spmap} do imply the shock acceleration in the discrete blobs (C and E1) in \obj\ \citep[see][]{2014AJ....147..143H}. However, it seems that the heavily knotty jets \citep[comparing with the jets in blazars,][]{2014AJ....147..143H} in \obj\ is inconsistent with the only mild flattening in the spectrum, i.e. the spectral distribution shows only plateaux than clear bumps \citep[see][]{2014AJ....147..143H}. This discrepancy may be indicative of an episodic scenario. Actually, the lacks of a bright radio core and the steep spectrum in the nascent ejecta C both support the episodic ejection scenario. In observations, there is also growing evidence to show that highly accreting AGNs tend to launch episodic jets \citep{2021MNRAS.508.1305Y,2022MNRAS.514.6215Y,2022ApJ...941...43Y,2022MNRAS.517.4959Y,2023MNRAS.520.5964Y}.

\subsection{On the physical interpretations of the complex jets}
By summing the radio flux density along the jet trajectory, i.e. only excluding the component S, the measured 5\,GHz jet luminosity of \obj\ is $\log{L_\mathrm{5GHz}}=38.547\pm0.003$ erg s$^{-1}$. Taking the X-ray luminosity of $\log{L_\mathrm{2-10keV}}=43.65$\,erg s$^{-1}$ measured in 2020 \citep{2021Natur.595..657W}, the radio to X-ray luminosity ratio of \obj\ is $L_\mathrm{R}/L_\mathrm{X}=10^{-5.102}$, suggesting the nature of the jet is the corona ejection from the accretion disk \citep{2009MNRAS.395.2183Y}, i.e. $L_\mathrm{R}/L_\mathrm{X}\sim10^{-5}$ \citep{2019NatAs...3..387P, 2020ApJ...904..200Y}, and consistent with the interpretation of the X-ray behavior \citep{2007MNRAS.377.1375G, 2017MNRAS.471.4436W}. Models and observational evidence suggest that episodic jets in both AGN and microquasar are ejected from the accretion disc corona \citep[see][and references therein]{2019ApJ...877..130S}.

The bilateral morphology and the linear size of $\sim$45 pc (the tapered 5\,GHz image in panel c of Figure~\ref{fig:vlbi} and Figure \ref{fig:ridgeline}) are reminiscent of \obj\ being a compact symmetric object \citep[CSO,][]{2021A&ARv..29....3O}. However, the lack of a spectral peak at GHz (Figure \ref{fig:si}) alternatively implies that it belongs to compact steep-spectrum (CSS) radio sources. Assuming minimum energy (approximately equipartition) conditions in synchrotron emission, we can estimate the magnetic field $B_\mathrm{min}$ of \obj\ in units of Gauss (G) through the formula \citep[e.g.][]{1980ARA&A..18..165M, 2022ApJ...934...26P}
\begin{equation}
B_\mathrm{min}\approx0.0152\left[\frac{a}{f_{rl}}\frac{(1+z)^{4-\alpha}}{\theta^3}\frac{S_i}{\nu^\alpha}\frac{\nu_2^{p+\alpha}-\nu_1^{p+\alpha}}{r(p+\alpha)}\right]^{2/7}~\mathrm{(G)},
\end{equation}
where $S_i$ (in mJy) is the integrated flux density of the source measured at frequency $\nu$ (in GHz) and angular size $\theta$ (in mas), $\alpha$ is the spectral index, $z$ is the redshift of the source, and $r$ is the comoving distance in Mpc. Here we take $p=0.5$, the overall spectral index $\alpha=-0.89\pm0.10$ from $\nu_1 \sim1$\,GHz to $\nu_2\sim15$\,GHz, a filling factor for the relativistic plasma $f_{rl} = 1$ and a relative contribution of the ions to the energy $a = 2$. By adopting the standard $\Lambda$CDM cosmology and using the cosmology calculator provided by NED \footnote{\url{https://www.astro.ucla.edu/~wright/CosmoCalc.html}}, $r=245.7\,\mathrm{Mpc}$ in this case. Here we use the integrated flux density of $2.636\pm0.282$\,mJy and the angular size of 50\,mas from the tapered VLBA 1.5\,GHz image, which yield an overall magnetic field $B_\mathrm{min}\approx10^{-3.6}\,\mathrm{G}$. The magnetic field is consistent with the typical value of e.g. CSOs with peaked-spectrum (PS, $B\sim10^{-3}\,\mathrm{G}$) and CSS ($B\sim10^{-4}\,\mathrm{G}$) radio sources \citep{1998PASP..110..493O}. On the other hand, the total radio power of \obj\ is only $10^{21.8}$\,W\,Hz$^{-1}$ (from the 5\,GHz EVN+$e$-MERLIN observational flux density of jets, see above), which is at least 1 order of magnitude lower than the typical sample of PS and CSS sources \citep{2021A&ARv..29....3O}. However, the radio power of \obj\ remains higher than our recently discovered CSO/PS in NGC\,4293 \citep[$\sim10^{20}$\,W\,Hz$^{-1}$, see][]{2022MNRAS.517.4959Y}, which hold the current lower limit record of radio power. Simultaneously, the magnetic field at component E1 can be estimated as $B_\mathrm{min}\approx10^{-2.9}\,\mathrm{G}$ by using the model-fitting results with either 1.5\,GHz VLBA or 5\,GHz EVN+$e$-MERLIN observational result.

As the promising spectral turnover of \obj\ must be below 1\,GHz, we can estimate the lowest allowed electron lifetime via the formula \citep{1998PASP..110..493O}
\begin{equation}
t\simeq8.22\times\frac{B^{1/2}}{B^2+B_{R}^2}\times{\left((1+z)\nu_p\right)}^{-1/2}~\mathrm{(yr)},
\end{equation}
where $B$ is the magnetic field in $\mathrm{G}$, $B_\mathrm{R}\simeq4(1+z)^2\times10^{-6}\,\mathrm{G}$ is the equivalent magnetic field of the microwave background, and $\nu_p$ is the break frequency in GHz. For \obj, using values estimated above, i.e. $B\approx10^{-3.6}\,\mathrm{G}$ and $\nu_p<1\,\mathrm{GHz}$, we find the electron lifetime is substantially $>10^{6.3}$\,years. Again, the estimated electron lifetime of component E1 is $>10^{5.2}$\,years. Taking component E1 as a reference (15\,mas away from the core) and assuming a typical knot advance speed of $0.1\,c$ for PS/CSS \citep{2021A&ARv..29....3O}, the ejection time scale of component E1 is only $\sim550$\,years and smaller than its electron lifetime above, therefore it suggests the knot advance speed of \obj\ should be substantially slower than $0.1\,c$. In stellar mass XRBs, discrete radio blobs are produced in the `very high state' or `super-Eddington state' at a time scale spanning from a few days to one year \citep[to the best of our knowledge, see][]{1989ApJ...347..448M, 1994Natur.371...46M, 1995Natur.375..464H, 1999MNRAS.304..865F, 2007MNRAS.375L..11T, 2011MNRAS.415L..59J, 2012MNRAS.421..468M, 2019Natur.569..374M, 2020NatAs...4..697B}. The longer ejection time scale in \obj\ seems consistent with its more massive nucleus than XRBs, i.e. the ejection time scale can be scaled through the mass of accretors, which supports the scale-invariant ejection process in both XRBs and AGNs.

Component S in the VLBA 1.5\,GHz image is a real structure (panel b of Figure \ref{fig:vlbi}) as it is identified in the 1.5\,GHz tapered map (panel (a) of Figure \ref{fig:vlbi}). The 5\,GHz EVN image fails to detect component S possibly due to the loss of large-scale and diffuse emission. The radio emission at component S is likely responsible for the flux density deficit in 5\,GHz (see supplementary Figure \ref{fig:fs} in Appendix \ref{apd:tab}). Furthermore, the southern bump in the VLA image (Figure \ref{fig:vla}) mimics component S, which requires further identification. Interestingly, component S clearly deviates from the jet trajectory, because the jet is along the East-West direction and extends up to 0.5\,arcsec scale (see Figure \ref{fig:vla}). Here we interpret the bending of component S as the result of jet-medium interactions. Given that the jet direction (extending up to $\sim0.5$\,parsec) is nearly aligned with the kpc-scale molecular disk \citep{2019ApJ...887...24T, 2020ApJ...899..112S} in \obj, it's possible to support a jet-disk interaction. On the other hand, \obj\ hosts strong multi-scale wide-angle outflows: an ultrafast wind-like outflow with a velocity of $>0.25\,c$ obtained from the fitting of the iron K-line profile \citep{2019ApJ...884...80R}, an ionized ultraviolet gas outflow with a velocity of $1870 {\rm \,km\,s}^{-1}$ \citep{1997ApJ...489..656L} and X-ray outflow with velocities of $\sim2000 {\rm \,km\,s}^{-1}$ \citep{2007MNRAS.378..873C, 2018MNRAS.480.2334S}, and a neutral gas outflow with a velocity of $45 {\rm \,km\,s}^{-1}$ \citep{2017ApJ...850...40R}. Given that the wide-angle outflows tend to be perpendicular to the kpc-scale molecular disk, for example, the neutral gas outflow \citep[see][for their Figure 13]{2017ApJ...850...40R} of \obj\ is along the North-South direction, therefore the component S can be a result of jet-wind collision as well.

\section{Summary}\label{sec:conclusion}
In summary, AGNs with near or super-Eddington accretion rates are often discussed as a scaled-up version of the stellar-mass black hole in the `very high state' or `ultraluminous state'. The observational evidence of episodic jets in \obj\ indicates that the analogy between AGNs and XRBs might also hold in this extreme state. Our observations imply that near or super-Eddington and extremely radio-quiet AGNs can also launch short-lived, small-scale, and weak jets. In the Supplementary section, we present further analysis of the radio emission in \obj, which sheds light on the long-standing question about the origin of radio emission in radio-quiet AGNs. Such a population is important for building our understanding of jet-disc coupling in near or super-Eddington states. There is only one Galactic microquasar (SS\,433) that exhibits long time-scale, super-Eddington behavior and quasi-continuous ejection \citep{2004ASPRv..12....1F}.  Only a few XRBs evolve into a near/super-Eddington state and are associated with episodic jets \citep{2002A&A...391.1013R, 2004ASPRv..12....1F, 2004MNRAS.349..393D, 2019Natur.569..374M}, but this phase is short-lived. Finally, the radio emission from a handful of (extragalactic) ultraluminous X-ray sources is too weak to be detected \citep{2017ARA&A..55..303K}.  The time scale of the super-Eddington state in AGNs is longer than the canonical `very high state' in XRBs according to the scaling relation, and the radio luminosity of the super-Eddington state in AGNs is higher than that of stellar-mass black holes according to the relation $L_\mathrm{R}/L_\mathrm{X}=10^{-5}$ \citep{2019NatAs...3..387P, 2020ApJ...904..200Y}, which is essential for testing the generic model of jet-disc coupling for all near/super-Eddington systems. Our findings here may indicate common features for high-Eddington AGNs, because there are similar knots/discrete radio blobs in jets and corresponding transverse/bending structures in other sources, e.g. NGC\,4051 \citep{2009ApJ...706L.260G} with Eddington ratio 0.2 \citep{2021ApJ...913....3Y} and Mrk\,335 \citep{2021MNRAS.508.1305Y} with Eddington ratio 0.48-3.11 \citep{2020ApJ...904..200Y}. Further observational and theoretical studies will be necessary to establish a comprehensive understanding of the outflows of near/super-Eddington AGNs, and the results obtained from this work may contribute to this goal.

\begin{acknowledgments}
This work is supported by the National Science Foundation of China (12103076, 11721303, 11991052), the National Key R\&D Programme of China (2016YFA0400702, 2018YFA0404602, 2018YFA0404603). 
XLY thanks the support of the Shanghai Sailing Program (21YF1455300) and the China Postdoctoral Science Foundation (2021M693267). 
SY was supported by an Alexander von Humboldt Foundation Fellowship. 
JS acknowledges financial support from the Czech Science Foundation project No.19-05599Y. 
MFG is supported by the Shanghai Pilot Program for Basic Research–Chinese Academy of Science, Shanghai Branch (JCYJ-SHFY-2021-013), the National SKA Program of China (Grant No. 2022SKA0120102), and the science research grants from the China Manned Space Project with NO. CMSCSST-2021-A06.
% Scientific data
Scientific results from data presented in this publication are derived from the EVN project EY037 and the VLBA project BY145.
% EVN
The European VLBI Network (EVN) is a joint facility of independent European, African, Asian, and North American radio astronomy institutes.
% e-MERLIN
$e$-MERLIN is a National Facility operated by the University of Manchester at Jodrell Bank Observatory on behalf of STFC.
% NRAO VLBA
The National Radio Astronomy Observatory is a facility of the National Science Foundation operated under cooperative agreement by Associated Universities, Inc.
% Gaia2
This work has made use of data from the European Space Agency (ESA) mission {\it Gaia} (\url{https://www.cosmos.esa.int/gaia}), processed by the {\it Gaia} Data Processing and Analysis Consortium (DPAC, \url{https://www.cosmos.esa.int/web/gaia/dpac/consortium}). Funding for the DPAC has been provided by national institutions, in particular the institutions participating in the {\it Gaia} Multilateral Agreement.
% DiFX
% This work made use of the Swinburne University of Technology software correlator, developed as part of the Australian Major National Research Facilities Programme and operated under licence. This work made use of data supplied by the UK Swift Science Data Centre at the University of Leicester.
\end{acknowledgments}


%\vspace{5mm}
%\facilities{HST(STIS), Swift(XRT and UVOT), AAVSO, CTIO:1.3m, CTIO:1.5m,CXO}


% \software{astropy \citep{2013A&A...558A..33A,2018AJ....156..123A},  
%          Cloudy \citep{2013RMxAA..49..137F}, 
%          Source Extractor \citep{1996A&AS..117..393B}
%          }



\appendix

\section{Large-scale radio emission} \label{apd:large}
Figure \ref{fig:si} shows \obj\ has a power-law radio spectrum over the entire observed frequency range. This indicates the dominance of synchrotron radiation, and it is not significantly affected by the size of collection areas (see supplementary Figure \ref{fig:fs} in Appendix \ref{apd:tab}) where the intercept between the L-band and C-band lines can be regarded as the spectral index. The overall spectral index is $-0.89\pm0.10$ and there is no significant decrease at higher frequencies, indicating a continuous replenishment of fresh electrons \citep{2017NatAs...1..596M}. Interestingly, the spectral index at the jet edge farthest from the core is $\sim-1.8$ (Figure \ref{fig:spmap}). This is inconsistent with the overall spectral index estimated for the larger area but suggests a non-jet origin because the spectral index decreases along the jet trajectory. The large-scale flux density is dominated by diffuse radio emission with only a fraction coming from the (parsec-scale) core region (only account $\sim30\%$ and $\sim47\%$ of radio emission from $60$\,kpc and $\sim1.54$\,kpc scale region at 1.5\,GHz, respectively). The distribution of radio flux density can be fitted as $S_L=(0.85\pm0.18)r^{0.212\pm0.020}$ and $S_C=(0.88\pm0.02)r^{0.131\pm0.005}$, where $S_L$ and $S_C$ is the L and C-band flux density in mJy, and $r$ is the angular size in mas (see supplementary Figure \ref{fig:fs} in Appendix \ref{apd:tab}). The VLBA 1.5\,GHz emission satisfies the flux density versus collection area distribution, while the 5\,GHz flux density from our EVN+$e$-MERLIN observation is underestimated due to the loss of large-scale emissions.

Both star-forming activities and relativistic winds can produce large-scale radio emission \citep{2019NatAs...3..387P}. Here the star-forming activities are preferentially referred to as supernovae or supernova remnants due to the power-law spectrum. Assuming all of the radio emission is from star-forming activities, we can estimate the star formation rate (SFR) from radio emission by using the SFR-radio relation \citep[formula 3 in][]{2020ApJ...904..200Y}. The largest SFR can be obtained from the datasets: NVSS at $1.4$\,GHz, AE0022 at $1.4$\,GHz and $4.86$\,GHz, and AA0048 at $14.94$\,GHz. These yield a SFR of $\sim20\,M_\odot\,yr^{-1}$, which is similar to other estimates \citep{2021ApJ...908..231M} of $\sim26\,M_\odot\,yr^{-1}$, suggesting the large scale radio emission can be entirely due to star-forming activities. Whilst the SFR-radio relation is crude and we can not fully rule out the contribution of wind-like outflow, the radio-emitting wind at large scales (a few kiloparsec scales) is negligible. In addition, the radio-emitting wind is still possible at the intermediate scale (tens of parsec scale), as there are no compact supernovae and supernova remnants detected in our VLBI and $e$-MERLIN observations \citep[e.g.][]{2008MNRAS.391.1384F}.

\section{Tabulated data and supplementary images}\label{apd:tab}

\begin{deluxetable*}{cccccc}
\tablecaption{Statistical parameters for different image deconvolution algorithms. \label{tab:sta}}
\tablewidth{0pt}
\tablehead{
\colhead{Algorithm} &\colhead{$\chi^2_{r}$}  &\colhead{d.o.f.} & \colhead{$\sigma_r$}          & \colhead{$\sigma_{rms}$} & \colhead{$f_{loss}$} \\
\colhead{}          &\colhead{}              &\colhead{}    & \colhead{(mJy\,beam$^{-1}$)}  & \colhead{(mJy\,beam$^{-1}$)}
}
\decimalcolnumbers
\startdata
\multicolumn{5}{c}{L-band} \\
\hline
Gaussian Model-fit    &$ 1.00524  $&$  3011564 $&$  0.03125     $&$  0.02735    $&$  0.14    $\\
%Delta Model-fit       &$ 1.00118  $&$   $&$  0.03390     $&$  0.02759    $&$  0.22    $\\
CLEAN                 &$ 1.00117  $&$ 3011221  $&$  0.03012     $&$  0.02735    $&$  0.10    $\\
\hline
\multicolumn{5}{c}{C-band} \\
\hline
Gaussian Model-fit    &$ 1.000489  $&$ 1418840 $&$  0.007362   $&$  0.005674    $&$  0.29    $\\
%Delta Model-fit       &$ 1.000495  $&$  $&$  0.008505   $&$  0.005669    $&$  0.50    $\\
CLEAN                 &$ 1.000076  $&$ 1418485 $&$  0.007537   $&$  0.005672    $&$  0.32    $\\
\enddata
\tablecomments{Column 1: Algorithm for deconvolution; Column 2: reduced Chi-squared value; Column 3: degrees of freedom; Column 4: rms noise in the residual image after subtraction of models; Column 5: off-source rms noise in the map; Column 6: $\frac{\sigma_r-\sigma_{rms}}{\sigma_{rms}}$, as a representative of the unrecovered flux density.}
\end{deluxetable*}

\begin{deluxetable*}{ccccccccccc}
\tablecaption{Summary of historical observations and results for \obj. \label{tab:archive}}
\tablewidth{0pt}
\tablehead{
\colhead{Telescope} &\colhead{$\nu$} & \colhead{Obs.\,ID}  & \colhead{Date} & \colhead{TOS} & \colhead{BW} & \colhead{$\theta_{maj}$} & \colhead{$\theta_{min}$} & \colhead{P.A.} & \colhead{S$_i$} & \colhead{S$_p$}\\
\colhead{}         &\colhead{(GHz)} &  \colhead{}    & \colhead{} & \colhead{(min)}    & \colhead{(MHz)}  & \colhead{(arcsec)}   & \colhead{(arcsec)}  & \colhead{(degree)} & \colhead{(mJy)} & \colhead{(mJy\,beam$^{-1}$)}
}
%\decimalcolnumbers
\startdata
VLA-D       &$  1.49 $& AE0022$^\dagger$         &    1983-07-04  &$   2.7     $&$ 100 $&$  52.7   $&$ 44.8 $&$ -14.9 $&$  8.40\pm1.40  $&$ 8.40\pm1.17 $\\
VLA-D       &$  1.40 $& NVSS$^{*,\diamond}$      &    1993-11-15  &$   0.4     $&$ 50  $&$  46     $&$ 46   $&$ 0     $&$  8.80\pm1.00  $&$ 7.60\pm0.46 $\\
VLA-D       &$  1.40 $& AV0226                    &    1996-09-17  &$   238     $&$ 6.2 $&$  59.2   $&$ 53.4 $&$ 25.0  $&$  8.43\pm0.31  $&$ 8.37\pm0.30 $\\
VLA-D       &$  1.40 $& AV0226                    &    1996-09-28  &$   237.5   $&$ 6.2 $&$  59.6   $&$ 53.5 $&$ 22.5  $&$  8.25\pm0.32  $&$ 8.27\pm0.30 $\\
VLA-C       &$  1.49 $& AB0417$^\star$           &    1986-11-08  &$   9.5     $&$ 100 $&$  18.9   $&$ 14.9 $&$ -35.1 $&$  6.68\pm0.67  $&$ 6.12\pm0.57 $\\
VLA-C       &$  1.34 $& AL0417                    &    1997-07-18  &$   190.8   $&$ 12.5$&$  10.2   $&$ 8.83 $&$ 5.19  $&$  6.45\pm0.48  $&$ 6.51\pm0.39 $\\
VLA-C       &$  1.34 $& AL0417                    &    1997-07-24  &$   195     $&$ 12.5$&$  11.1   $&$ 8.34 $&$ -33.2 $&$  6.49\pm0.41  $&$ 6.17\pm0.40 $\\
VLA-B       &$  1.49 $& AG0173                    &    1985-04-22  &$   85      $&$ 100 $&$  5.94   $&$ 4.69 $&$  52.4 $&$  5.13\pm0.73  $&$ 4.19\pm0.43 $\\
VLA-B       &$  1.40 $& FIRST$^{*,\circ}$        &    2011-03-26  &$   1       $&$ 256 $&$  4.3    $&$ 4.3  $&$  0    $&$  5.30\pm0.24        $&$ 5.05\pm0.15 $\\
VLA-A       &$  1.42 $& AL0502                    &    1999-08-26  &$   5.5     $&$ 100 $&$  1.36   $&$ 1.27 $&$  10.9 $&$  5.18\pm0.27  $&$ 6.77\pm0.24 $\\
VLA-A       &$  1.42 $& AK0406                    &    1995-07-20  &$   6.1     $&$ 100 $&$  1.68   $&$ 1.41 $&$ -0.62 $&$  5.17\pm0.19  $&$ 4.89\pm0.16 $\\
VLA-A       &$  1.42 $& AC0624                    &    2002-05-02  &$  10.5     $&$ 100 $&$  1.66   $&$ 1.36 $&$ -5.28 $&$  5.28\pm0.16  $&$ 4.90\pm0.13 $\\
\hline
Arecibo     &$  2.38 $& NG$^{*,\divideontimes}$  &    1975-08-01  &$           $&$ 16  $&$  162    $&$ 162  $&$  0    $&$  4.00\pm4.00  $&$             $\\
VLA-BnA     &$  2.98 $& VLASS1.1$^{*,\Join}$     &    2017-10-08  &$   0.1     $&$ 2048$&$  3.06   $&$ 2.28 $&$  52.7 $&$  3.71\pm0.33  $&$ 2.87\pm0.11 $\\
VLA-BnA     &$  2.98 $& VLASS2.1$^{*,\Join}$     &    2020-07-16  &$   0.1     $&$ 2048$&$  2.68   $&$ 2.37 $&$  36.6 $&$  3.63\pm0.35  $&$ 2.72\pm0.17 $\\
\hline
VLA-D       &$  4.86 $& AE0022                    &    1983-07-04  &$   12.7    $&$ 100 $&$  12.9   $&$ 11.1 $&$ -10.5 $&$  3.16\pm0.18  $&$ 2.27\pm0.10 $\\
VLA-C       &$  4.86 $& AK0298                    &    1992-05-18  &$   27.5    $&$ 100 $&$  9.83   $&$ 4.90 $&$  55.9 $&$  2.80\pm0.11  $&$ 1.33\pm0.04 $\\
VLA-C       &$  4.86 $& AB0417$^\star$           &    1986-11-08  &$   9       $&$ 100 $&$  5.56   $&$ 4.72 $&$ -33.9 $&$  2.46\pm0.15  $&$ 2.47\pm0.08 $\\
VLA-B       &$  4.86 $& AL0454                    &    1998-09-02  &$   8.6     $&$ 100 $&$  1.51   $&$ 1.38 $&$  7.61 $&$  2.20\pm0.09  $&$ 2.13\pm0.05 $\\
VLA-A       &$  4.86 $& AK0096$^\S$              &    1983-11-24  &$   25.1    $&$ 100 $&$  0.50   $&$ 0.42 $&$ -29.1 $&$  1.84\pm0.08        $&$ 1.79\pm0.04 $\\
VLA-A       &$  4.86 $& AK0406                    &    1995-07-20  &$   9.5     $&$ 100 $&$  0.49   $&$ 0.41 $&$ -2.14 $&$  2.18\pm0.06  $&$ 2.02\pm0.05 $\\
VLA-A       &$  4.86 $& AC0624                    &    2002-05-02  &$   9.8     $&$ 100 $&$  0.48   $&$ 0.41 $&$ 11.6  $&$  1.84\pm0.09  $&$ 1.85\pm0.05 $\\
MERLIN      &$  4.99 $& 96DECA                    &    1996-12-14  &$   558     $&$ 15  $&$  0.08   $&$ 0.05 $&$ 22.35 $&$  1.46\pm0.22  $&$ 1.43\pm0.11 $\\
MERLIN      &$  4.99 $& 97NOVC                    &    1997-11-06  &$   450     $&$ 15  $&$  0.08   $&$ 0.04 $&$ 20.04 $&$  1.51\pm0.52  $&$ 1.83\pm0.27 $\\
\hline
VLA-B       &$  8.46 $& AC0624                    &    2002-08-09  &$   9.8     $&$ 100 $&$  0.68   $&$ 0.65 $&$ 20.2  $&$  1.46\pm0.07  $&$ 1.15\pm0.03 $\\
VLA-A       &$  8.44 $& AP0212                    &    1991-06-25  &$   14.8    $&$ 100 $&$  0.26   $&$ 0.24 $&$ 59.7  $&$  0.94\pm0.10  $&$ 0.90\pm0.05 $\\
VLA-A       &$  8.44 $& AB0670                    &    1992-12-04  &$   119.3   $&$ 100 $&$  0.27   $&$ 0.26 $&$ -12.2 $&$  1.18\pm0.03  $&$ 1.09\pm0.01 $\\
\hline
VLA-C       &$ 14.94 $& AA0048$^\star$           &    1985-07-28  &$   14.5    $&$ 100 $&$  1.87   $&$ 1.49 $&$ -29.5 $&$  1.20\pm0.46  $&$ 1.20\pm0.26 $\\
\hline
OVRO-40     &$ 20.00 $& NG$^{*,\dagger}$         &    1983-07-07  &$           $&$ 400 $&$  90     $&$ 90   $&$  0    $&$  <1.90        $&$ <1.90       $\\
VLA-C       &$ 22.46 $& AA0048$^\star$           &    1985-07-28  &$   29      $&$ 100 $&$  1.13   $&$ 1.01 $&$ -48.9 $&$  <1.91        $&$ <1.91       $\\
VLA-A       &$ 22.46 $& AP0210                    &    1991-07-02  &$  18.5     $&$ 100 $&$  0.38   $&$ 0.09 $&$ -56.8 $&$  <1.76        $&$ <1.76       $\\
\hline
ALMA        &$ 108.64$& NG$^{*,\ddagger}$         &    2016-08-01  &$   4.0     $&$     $&$  0.57   $&$ 0.46 $&$       $&$  0.49\pm0.07  $&$             $\\
\enddata
\tablecomments{Column 1: telescope; Column 2: frequency; Column 3: project ID and references for datasets publication (NG: Not given; *: Original data is not available, the result is from literature); Column 4: observing date; Column 5: time-on-source; Column 6: observing bandwidth; Column 7 - 9: beam major axis, minor axis, and position angle; Column 10: integrated flux density; Column 11: peak flux density.}
\tablecomments{References: $\dagger$: \citet{1987ApJ...313..651E}; $\diamond$: \citet{1998AJ....115.1693C}; $\star$: \citet{1989ApJS...70..257B} and \citet{1996AJ....111.1431B}; $\circ$: \citet{1997ApJ...475..479W}; $\divideontimes$:\citet{1978ApJS...36...53D}; $\Join$: \citet{2020PASP..132c5001L}; $\S$: \citet{1989AJ.....98.1195K}; $\ddagger$: \citet{2019ApJ...887...24T}.}
\end{deluxetable*}

\begin{deluxetable*}{ccccc}
\tablecaption{Coordinates of the calibrator J0056$+$1341 and the target \obj. \label{tab:coord}}
\tablewidth{0pt}
\tablehead{
\colhead{Observed feature} &\colhead{$\alpha$ (J2000)}  &\colhead{$\Delta\alpha$} & \colhead{$\delta$ (J2000)}   & \colhead{$\Delta\delta$} \\
\colhead{}                 &\colhead{(h m s)}           &\colhead{(s)}            & \colhead{($^\circ$ $^\prime$ $^{\prime\prime}$)}        & \colhead{($^{\prime\prime}$)}
}
\decimalcolnumbers
\startdata
J0056$+$1341 \\
\hline
J1 at 1.5\,GHz    & 00:56:14.816082  &  0.000002     &  $+$13:41:15.76430     &  0.00014     \\
J1 at 5\,GHz      & 00:56:14.8159647  &  0.0000008   &  $+$13:41:15.767536    &  0.000039    \\
The brightest feature at 4.34\,GHz (Astrogeo)    & 00:56:14.8161000  &  0.0000002     &  $+$13:41:15.754988    &  0.000009    \\
The brightest feature at 7.62\,GHz (Astrogeo)    & 00:56:14.8160522  &  0.0000001     &  $+$13:41:15.755279    &  0.000005    \\
\hline
\obj \\
\hline
$Gaia$ DR2                                   & 00:53:34.933288  &  0.000012   &   $+$12:41:35.9308    &  0.00017    \\
The brightest feature at VLBA 1.5\,GHz (E1)  & 00:53:34.934176  &  0.000015   &   $+$12:41:35.9267    &  0.00022    \\
The brightest feature at EVN  5\,GHz (E1)    & 00:53:34.934112  &  0.000012   &   $+$12:41:35.9275    &  0.00017    \\
\enddata
\tablecomments{The position errors presented here are only from random noise.}
\end{deluxetable*}

\begin{deluxetable*}{cccccccc}
\tablecaption{Model-fitting results of the radio components detected in \obj\ with the VLBA 1.5\,GHz and EVN+$e$-MERLIN 5\,GHz observations. \label{tab:vlbi}}
\tablewidth{0pt}
\tablehead{
\colhead{Component} &\colhead{$\nu$} & \colhead{R.A.Off}  & \colhead{DEC.Off} & \colhead{$S_i$} & \colhead{Size} & \colhead{$\theta_{lim,maj}\times\theta_{lim,min}$} & \colhead{$\log{T_B}$} \\
\colhead{}         &\colhead{(GHz)} &  \colhead{(mas, J2000)}    & \colhead{(mas, J2000)} & \colhead{($\mathrm{mJy}$)}    & \colhead{(mas)}  & \colhead{(mas)} & \colhead{($\mathrm{K}$)}
}
\decimalcolnumbers
\startdata
C           &$ 1.548     $&$     0.55\pm0.38  $&$   -1.31\pm0.76  $&$  0.42\pm0.05  $&$  4.5  $&$ 1.60\times0.64 $&$  7.22  $\\
E1           &$ 1.548     $&$    11.67\pm0.26   $&$  -3.31\pm0.50  $&$  1.28\pm0.08  $&$  5.1  $&$  0.92\times0.36 $&$  7.57  $\\
E2           &$ 1.548     $&$    18.60\pm0.52  $&$   -6.88\pm0.66  $&$  0.73\pm0.07  $&$  7.2  $&$  1.21\times0.48 $&$  7.05  $\\
S            &$ 1.548     $&$    -2.12\pm1.51  $&$   -17.61\pm2.24 $&$  0.74\pm0.10  $&$ 20.0 $&$  3.06\times1.22 $&$  6.17  $\\
E3           &$ 1.548     $&$    26.25\pm0.71  $&$   -2.65\pm1.56  $&$  0.12\pm0.05  $&$  2.4  $&$  1.20\times0.48 $&$  7.19  $\\
\hline
C           &$ 4.926     $&$    1.26\pm0.21   $&$  -0.77\pm0.46   $&$   0.15\pm0.01  $&$ 0.7  $&$  0.62\times0.22 $&$  7.29   $\\
E1           &$ 4.926     $&$   12.05\pm0.27   $&$  -3.29\pm0.47   $&$   0.47\pm0.05  $&$ 6.2  $&$  0.54\times0.19 $&$  5.97   $\\
E2           &$ 4.926     $&$   20.95\pm0.98   $&$  -10.04\pm1.00  $&$   0.15\pm0.03  $&$ 9.0  $&$  0.31\times0.11 $&$  5.17   $\\
W1           &$ 4.926     $&$   -8.85\pm0.68   $&$   2.31\pm0.82   $&$   0.11\pm0.02  $&$ 7.1  $&$  0.86\times0.30 $&$  5.26   $\\
\enddata

\tablecomments{Column 1: component name; Column 2: frequency; Column 3-4: right ascension and declination offset correspond to the $Gaia$ DR2 position; Column 5: integrated flux density; Column 6: the angular size of components from a Gaussian model-fit; Column 7: resolution limit along the major and minor axis direction of the synthesized beam; Column 8: lower limit of the radio brightness temperature.}
\end{deluxetable*}

\setcounter{figure}{0}

\begin{figure*}
\centering
\includegraphics[scale=0.6]{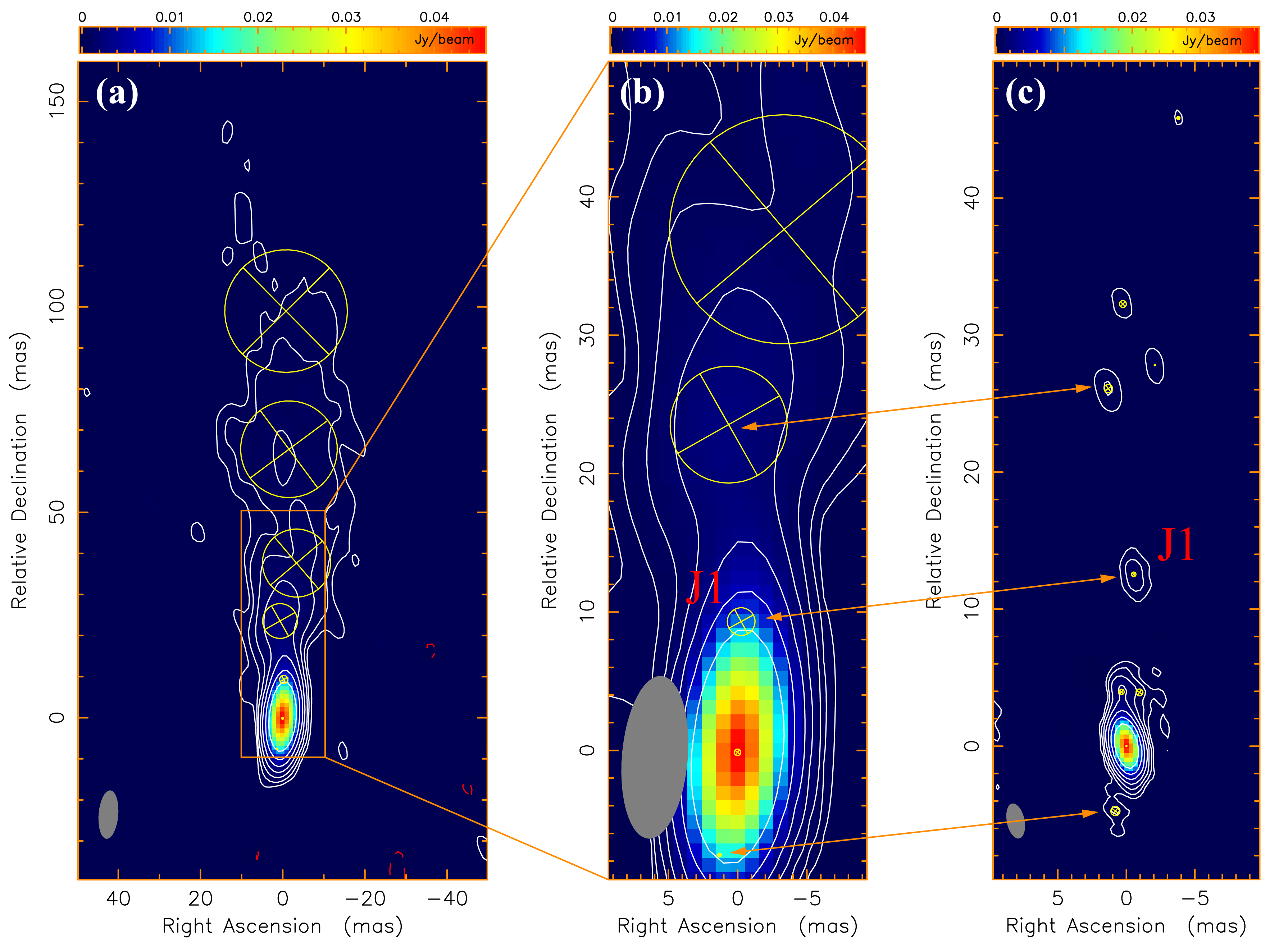}
\caption{\textbf{Model-fitting images of the phase calibrator J0056$+$1341 at $1.5$ GHz (panel a and b) and $5$\,GHz (panel c)}. The images are produced using a two-dimensional Gaussian model fit with natural weights. The contours are plotted as $3\sigma\times(-1,1,2,4,8,\dots)$, where $\sigma$ is the root mean square (rms) noise. The white solid curves represent positive values and the red dashed curves represent negative values. The rms noise is $0.2$\,mJy/beam for both $1.5$ and $5$\,GHz images. The model-fitting components are superimposed as yellow circles. The grey ellipses in the bottom left corner of each panel represent the full width at half-maximum (FWHM) of the restoring beam. The grey lines between panels c and b indicate the corresponding components without the core-shift effect, i.e. the optically thin components.}\label{fig:j0056}
\end{figure*}

\begin{figure*}
\centering
\includegraphics[scale=0.8]{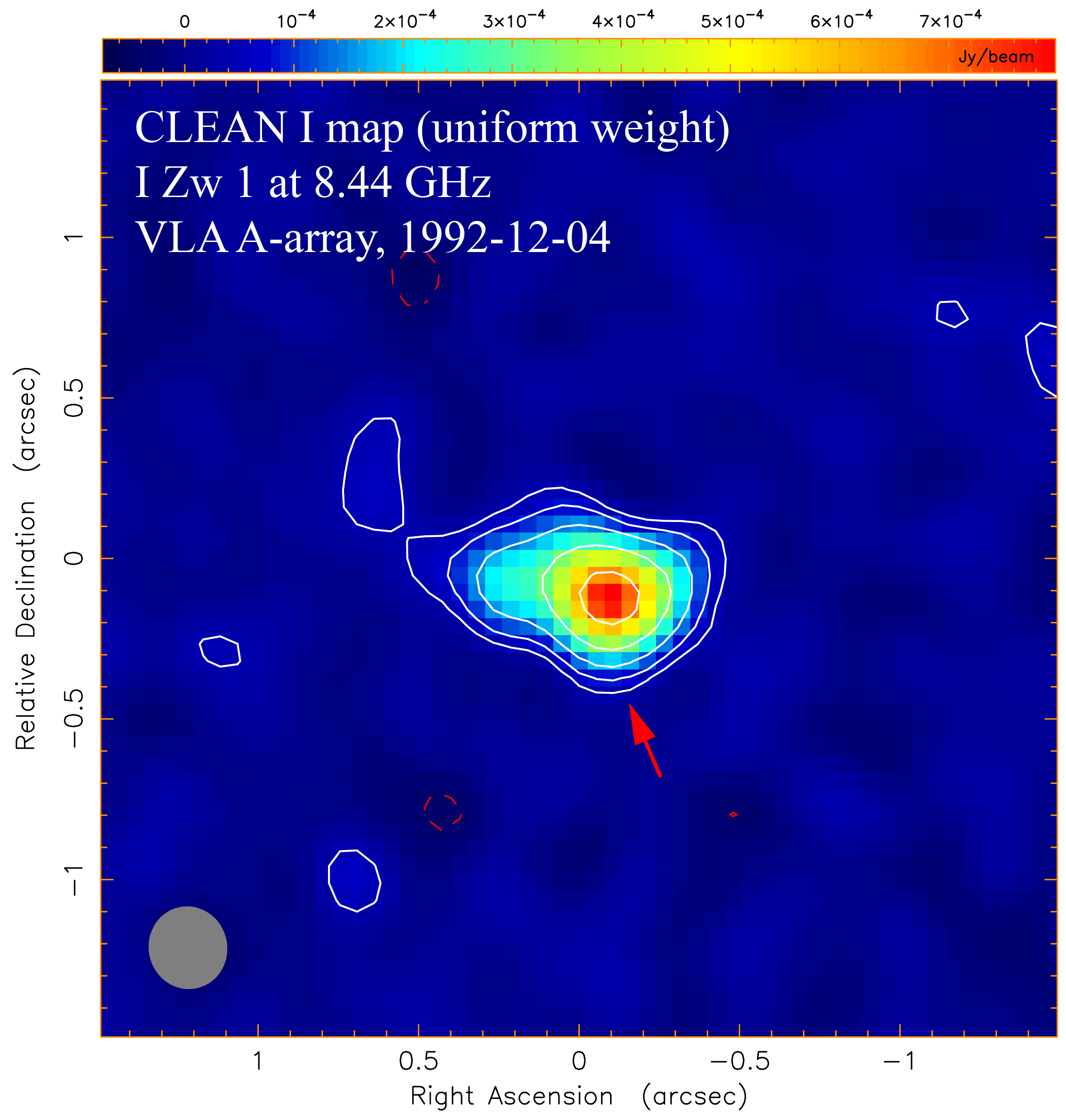}
\caption{\textbf{VLA A-array 8.4\,GHz image of \obj.} $Gaia$ DR2 position of \obj\ is set as the map center. The image peak is $0.798$\,mJy\,beam$^{-1}$. The contours are at 3$\sigma\times (-1, 1, 2, 2, 4, 8, \dots)$ and $1\sigma=0.038$\,mJy\,beam$^{-1}$, where positive contours are white and negative ones are red dashed. FWHM of restoring beam is $0.258\times0.245$\,arcsec at 12.7$^\circ$ and displayed as grey ellipses in the lower-left corner. The red arrow marks a possible southern bump.}\label{fig:vla}
\end{figure*}

\begin{figure*}
\centering
\includegraphics[scale=0.6]{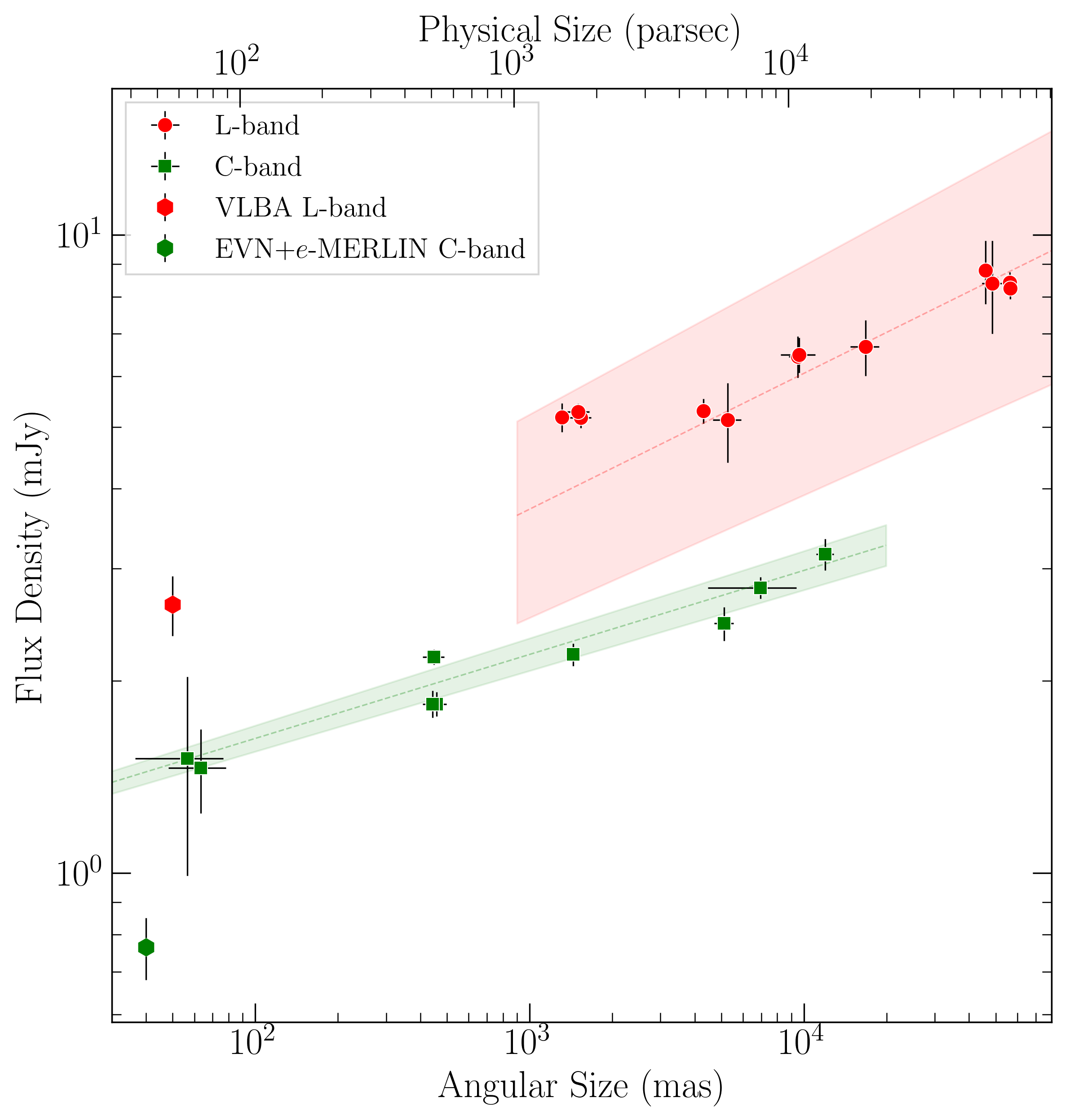}
\caption{\textbf{The radio flux density of \obj\ over a collection area range from $\sim$0.04 to $\sim$50\,arcsec.} The integrated radio flux densities and uncertainties of \obj\ in L and C-band are shown (Table \ref{tab:archive}). As \obj\ is not resolved in the given observations, the synthesized beams are taken to represent the collection area. The dashed lines and belts show power-law fittings and 95\% confidence intervals, respectively, in the L (red) and C-band (green).}\label{fig:fs}
\end{figure*}

%\section{Rotating tables} \label{sec:rotate}

%The process of rotating tables into landscape mode is slightly different in
%\aastex v6.31. Instead of the {\tt\string\rotate} command, a new environment
%has been created to handle this task. To place a single page table in a
%landscape mode start the table portion with
%{\tt\string\begin\{rotatetable\}} and end with
%{\tt\string\end\{rotatetable\}}.

%Tables that exceed a print page take a slightly different environment since
%both rotation and long table printing are required. In these cases start
%with {\tt\string\begin\{longrotatetable\}} and end with
%{\tt\string\end\{longrotatetable\}}. Table \ref{chartable} is an
%example of a multi-page, rotated table. The {\tt\string\movetabledown}
%command can be used to help center extremely wide, landscape tables. The
%command {\tt\string\movetabledown=1in} will move any rotated table down 1
%inch. 

\clearpage
\bibliography{izw1_apjl}{}
\bibliographystyle{aasjournal}



\end{document}